\documentclass[12pt]{article}
\usepackage[dvips]{color,graphicx}
\usepackage{hyperref}
\usepackage{amsmath,amssymb}
\usepackage{epsfig}
\usepackage[english]{babel}
\usepackage{pgf,pgfarrows,pgfnodes,pgfautomata,pgfheaps}
\usepackage{amsfonts,amsmath,amsthm,amssymb,enumerate,array,amssymb}
\usepackage[latin1]{inputenc}
\usepackage{bm}
\usepackage{fixmath}
\usepackage{amsbsy}
\usepackage{graphicx}
\usepackage{capt-of}
\usepackage{dcolumn}
\usepackage{yfonts}

\newcommand{\be}{\begin{equation}}
\newcommand{\ee}{\end{equation}}
\newcommand{\bea}{\begin{eqnarray}}
\newcommand{\eea}{\end{eqnarray}}
\newcommand{\pxi}{\frac{\partial}{\partial x^i}}

\begin{document}

\begin{titlepage}

\begin{flushright}
UTTG-08-12\\
TCC-010-12
\end{flushright}
\vspace{0.3cm}
\begin{center} \Large \bf Kinetic Theory of Collisionless Self-Gravitating Gases:\\
II. Relativistic Corrections in Galactic Dynamics
\end{center}
\vspace{0.1cm}
\begin{center}
J. Ramos-Caro$^{\star}$\footnote{javier@ime.unicamp.br},
C. A. Ag\'on$^{\dagger}$\footnote{cesar.agon@nucleares.unam.mx}
and J. F. Pedraza$^{\natural}$\footnote{jpedraza@physics.utexas.edu}

\vspace{0.2cm}
$^{\star}$
Departamento de Matem\'atica Aplicada, IMECC, Universidade Estadual de Campinas,\\
Campinas, S\~ao Paulo 13083-859, Brazil\\
 \vspace{0.2cm}
$^{\dagger}\,$Instituto de Ciencias Nucleares, Universidad Nacional Aut\'onoma de
M\'exico,\\ Apartado Postal 70-543, M\'exico D.F. 04510, M\'exico\\
 \vspace{0.2cm}
$^{\natural}\,$Theory Group, Department of Physics and Texas Cosmology Center, \\
University of Texas, 1 University Station C1608, Austin, TX 78712, USA\\
\vspace{0.2cm}
\end{center}

\begin{center}
{\bf Abstract}
\end{center}
\noindent
In this paper we study the kinetic theory of many-particle astrophysical
systems imposing axial symmetry and extending our previous analysis in Phys. Rev. D \textbf{83}, 123007 (2011).
Starting from a Newtonian model describing a collisionless self-gravitating gas, we develop a framework to
include systematically the first general relativistic corrections to the matter distribution and gravitational
potentials for general stationary systems. Then, we use our method to obtain particular solutions for the case
of the Morgan \& Morgan disks. The models obtained are fully analytical and correspond to the post-Newtonian
generalizations of classical ones. We explore some properties of the models in order to estimate the importance of
post-Newtonian corrections and we find that, contrary to the expectations, the main modifications appear far from the galaxy cores. As a by-product of this investigation we derive the corrected version of the tensor virial theorem. For stationary systems we recover the same result as in the Newtonian
theory. However, for time dependent backgrounds we find that there is an extra piece that contributes to the variation of the inertia tensor.
\vspace{0.2in}
\smallskip
\end{titlepage}
\setcounter{footnote}{0}

\tableofcontents

\section{Introduction}\label{introduccion}

The dynamics and evolution of collisionless stellar ensembles is a subject of great interest in astrophysics since they
are the primary tool for comparisons of observations and theory in galactic dynamics \cite{BT}. Such systems are generally
composed by billions of stars, so it is neither
practical nor worthwhile to follow the orbit of each particle in the ensemble. Most testable predictions depend
only on the distribution function $F(\mathbf{x},\mathbf{v},t)$ (DF), a quantity that determines the probability of
finding a single star in a given phase-space volume $d^3\mathbf{x}d^3\mathbf{v}$ around the position $\mathbf{x}$ and
the velocity $\mathbf{v}$. The dynamics of the DF follows from the appropriate kinetic equation and it in turn determines
the statistical evolution of the system.

In the framework of the general theory of relativity (GR) it is assumed that the DF
satisfies the general relativistic version of the Fokker-Planck equation \cite{kand1,kand2,kand3}
or the collisionless Boltzmann equation (CBE) \cite{kremer,chacon}. The first one is devoted to systems
in which local gravitational encounters dominate in their evolution whereas the latter is
useful to study systems sufficiently smooth, so that they may be considered to be
collisionless \cite{BT}. One can actually consider systems in which a number of particle species can
collide and produce different species. This is how the formation of the light elements in the big bang
 nucleosynthesis is calculated (see \cite{nucleus} for a review).
However, in systems such as galaxies and galaxy clusters,
physical encounters between the stars are very rare, and the effect of gravitational collisions can be neglected
for times far longer than the age of the universe. Thus, for these systems, the CBE provides a very good approximation.

For a typical galaxy, the relaxation time, $t_{{\rm relax}}$, is
arbitrarily large in comparison with the crossing time, $t_{{\rm cross}}$. This means that
they can be approximated as a continuum rather than concentrated into nearly pointlike stars.
Now, it is commonly assumed that
the main contribution of the mass in a galaxy is concentrated in an axisymmetric flat
distribution \cite{BT}. For this reason the obtention of idealized thin disk models has been a problem of great astrophysical relevance.
In this case, the most straightforward way to construct a self-consistent model
is by means of finding the DF for a system with a known gravitational interactions
and matter distribution. Since the mass density of the system is defined by the
integration of the DF over the velocity, the problem of finding a DF is that
of solving an integral equation (see \cite{eddi,frick,lynden,hunt93,jiao,PRGfrac}
 and the references therein). At present we have at disposal a variety of self-consistent galaxy models:
  \cite{LiBe,mestel,T1,T2,HunterToomre,MYM,KAL,jiang00,jiangmos02,jiangoss06,GR,PRG}.

Now, even though for most systems under consideration Newtonian gravity is believed to be dominant, general
relativistic corrections might play an important role in their evolution.
As a matter of fact, in recent years it has been an increasing interest in the
incorporation of GR in the description of these systems, and to date we have a variety of fully relativistic galaxy models:
 \cite{Lemos1,Lemos2,Lemos3,gonzalez-letelier,gonzalez-letelier2,semerak-zacek1,semerak-zacek2,semerak-zacek3,lop},
among others. Perhaps the principal reason of including GR corrections in galactic dynamics,
is the hypothesis that it is possible to overcome the problem of the rotation curves predicted by the Newtonian theory.
While, some authors argue that by using GR the inclusion of a dark matter halo is
unnecessary at galactic scales (see for instance \cite{Cooperstock-Tieu,Pro-Cooperstock1,Pro-Cooperstock2,Pro-Cooperstock3,Pro-Cooperstock4}), several
publications have pointed out that this is not entirely true \cite{Anticoperstok1,Anticoperstok2,Anticoperstok3,Anticoperstok4,Anticoperstok5}.
In particular, the authors of \cite{grumiller} presented a model in which the percentage of dark matter needed to explain flat rotation
curves turns out to be $\sim\!30\%$ less than the required by the Newtonian theory\footnote{Strictly speaking, the aim of introducing a dark matter halo is not necessarily to reproduce a flat rotation curve but generate a curve that rises and then declines at particular paces \cite{salucci}.}.
It is important to point out that currently there are alternative approaches to GR which address the problem of rotation curves in spiral
galaxies, as for example the so-called $f(R)$ gravity (see \cite{capozziello}, for references).

Despite the fact that the relativistic contributions do not solve completely the problem of rotation curves in galaxies,
it seems that they do introduce significant corrections. Thus, in order to estimate the effects on the various observables we are interested in,
it would be nice to have a framework to include systematically general relativistic corrections to a given Newtonian model.
The post-Newtonian approximation is perfectly suited for this
purpose. The appropriate scheme that describes the effects of the first
corrections beyond the Newtonian theory, was first formulated in \cite{eins1, eins2, eins3} (see \cite{WB} for a textbook analysis) and it is
known as the first post-Newtonian (1PN) approximation. This approach holds if the particles in the system are moving nonrelativistically ($\bar{v}\ll c$) (as in the case of a star moving around a typical
galaxy) and gives the corrections up to order $\bar{v}^{2}/c^{2}$, where $\bar{v}$ is a typical velocity in the system and $c$ is
the speed of light. Currently, higher order PN approximations have been developed
because of the increasing interest around kinematics and associated emission of
gravitational waves by binary pulsars, neutron stars and black holes, with
promising candidates for detectors such as LIGO, VIRGO and GEO600 (see
\cite{futamase, bichak} for references).

Based in the above considerations, we recently started a general study of self-gravitating gases in the collisionless
regime and, as a first step, we derived a version of the CBE that accounts for the first general relativistic corrections \cite{pedraza-agon-ramos}.
With this tool in hand, we obtained the
1PN version of the Eddington's polytropes, starting from an ergodic DF
proportional to $E^{n}$. The purpose of this paper is twofold.
First, to implement a similar procedure in
the axially symmetric case in order to setup our general framework. And second,
to obtain a new set of self-consistent models starting from a Newtonian ``seed'' and study
the impact of relativistic corrections on the various observables.

The rest of the paper is organized as follows. In section \ref{sec:gen-frame} we present
a brief overview about the basics of the 1PN approximation, revisiting the field equations,
as well as the kinetic theory for arbitrary self-gravitating systems.
In section \ref{sec:1PN-self-grav-eq} we show the fundamental equations defining self-consistent models with post-Newtonian corrections. We start dealing with arbitrary systems but then we focus on discoidal
configurations with axial symmetry, in order to prepare the ground to construct 1PN galaxy models in section \ref{sec:analyitical-models}. Finally, we summarize the principal results in section \ref{Conclusions}.

\section{General Framework}\label{sec:gen-frame}

\subsection{The 1PN approximation}

The post-Newtonian approximation has been reviewed carefully in a number of references (see for example \cite{WB}).
However, we will include here the basic definitions and relations for completeness.

First off, note that in Newtonian mechanics the typical kinetic energy is roughly of the same order of magnitude as the typical potential energy, and thus
$$
\bar{v}^2\sim\,\phi,
$$
where $\bar{v}$ is the mean velocity in the system.
The idea is then to express all physical quantities in terms of a series expansion of the small parameter $\bar{v}/c\ll1$
and keep the leading order beyond the Newtonian theory. The first quantity to consider is the spacetime itself: any manifold
can be considered to be locally flat so, for particles that are moving nonrelativistically, we proceed to express the metric tensor as

\bea
g_{00}&=&-1 +\stackrel{2 \: \: \: \: }{g_{00}}
+\stackrel{4 \: \: \: \: }{g_{00}}+\cdots,\nonumber\\
g_{ij}&=&\delta_{ij}+ \stackrel{2 \: \: \: \: }{g_{ij}} +\stackrel{4 \: \: \: \: }{g_{ij}}+\cdots,\\
g_{0i}&=& \stackrel{1 \: \: \: \: }{g_{0i}} +\stackrel{3 \: \: \: \: }{g_{0i}} +\stackrel{5 \: \: \: \: }{g_{0i}}+\cdots,\nonumber
\eea
where the symbol $\stackrel{n \: \: \: \: }{g_{\mu\nu}}$ denotes the term in $g_{\mu\nu}$ of order $({\bar v}/c)^n$.
Odd powers of ${\bar v}/c$ appear in $g_{0i}$ because
these components must change sign under time-reversal transformation $t\rightarrow -t$.

It is natural to assume a similar expansion for the components of the energy momentum tensor.
From their interpretation as the energy density, momentum flux, and energy flux, we expect that
\bea
T^{00}&=&\stackrel{0 \: \: \: \: }{T^{00}} +\stackrel{2 \: \: \: \: }{T^{00}}+\cdots,\nonumber\\
T^{ij}&=&\stackrel{2 \: \: \: \: }{T^{ij}} +\stackrel{4 \: \: \: \: }{T^{ij}}+\cdots,\\
T^{0i}&=&\stackrel{1 \: \: \: \: }{T^{0i}}+\stackrel{3 \: \: \: \: }{T^{0i}} +\cdots.\nonumber
\eea
These expansions lead to a consistent solution of Einstein field equations.

Working in harmonic coordinates (i.e., coordinates such that $g^{\mu\nu}\Gamma^\lambda_{\mu\nu}=0$)
and to our order of approximation, the various components of the metric tensor can be expressed in terms of
the Newtonian potential $\phi$ and post-Newtonian potentials $\psi$ and $\xi_i$ as
\be
\begin{split}\label{potentials}
\stackrel{2 \: \: \: \: }{g_{00}}&=\,-2\phi/c^2,\\
\stackrel{4 \: \: \: \: }{g_{00}}&=\,-2(\phi^2+\psi)/c^4,\\
\stackrel{2 \: \: \: \: }{g_{ij}}&=\,-2\phi\delta_{ij}/c^2,\\
\stackrel{1 \: \: \: \: }{g_{0i}}&=\,0,\\
\stackrel{3 \: \: \: \: }{g_{0i}}&=\,\xi_i/c^3.
\end{split}
\ee
Thus, the Einstein equations reduce to
\bea
\nabla^2 \phi &=& 4\pi G \;\stackrel{0 \: \: \: \: }{T^{00}}, \label{ec_campo_pn1} \\
\nabla^2 \psi &=& 4\pi G c^{2}( \;\stackrel{2 \: \: \: \: }{T^{00}}
+ \;\stackrel{2 \: \: \: \: }{T^{ii}}) + \frac{\partial^2 \phi}{\partial t^2}, \label{ec_campo_pn2} \\
\nabla^2 \xi_i &=& 16 \pi G c \;\stackrel{1 \: \: \: \: }{T^{0i}}, \label{ec_campo_pn3}
\eea
along with the coordinate condition
\begin{equation}\label{condition-phi-xi}
    4\frac{\partial\phi}{\partial t}+\nabla\cdot\mbox{\boldmath$\xi$}=0.
\end{equation}

One can also consider the motion of test particles in a given background. For general potentials $\phi$, $\psi$, and $\xi_i$ one finds that
the free falling particle obeys the equation
\be
\frac{d \mathbf{v}}{d t}=-\nabla\phi-\frac{1}{c^{2}}\left[\nabla\left(2\phi^{2}+\psi\right)+\frac{\partial \mbox{\boldmath$\xi$}}{\partial t}-\mathbf{v}
\times(\nabla\times\mbox{\boldmath$\xi$})-3\mathbf{v}\frac{\partial\phi}{\partial t}-4\mathbf{v}(\mathbf{v}\cdot\nabla\phi)
+v^{2}\nabla\phi\right],\label{ecmovimiento1PN}
\ee
which partially resembles the mathematical structure of the Lorentz force experienced by a charged particle, with velocity $\mathbf{v}$,
in the presence of an electromagnetic field. Such law of
motion will determine, for instance, the rotation curve corresponding to a given galactic model (see Appendix \ref{Circular Velocity1PN} for details).

It is instructive to point out that the equations of motion (\ref{ecmovimiento1PN})
can be derived from the Lagrangian \cite{WB}
\begin{equation}\label{lag-wein}
\mathfrak{L}=\frac{v^2}{2}-
\phi-\frac{1}{c^2}\left(\frac{\phi^2}{2}+\frac{3\phi v^2}{2}-\frac{v^4}{8}+\psi-\mathbf{v}\cdot\mbox{\boldmath$\xi$}\right).
\end{equation}
For stationary spacetimes, the potentials are independent of time and the associated Hamiltonian
$\mathcal{H}=\sum_i\dot{x}_i\partial\mathfrak{L}/\partial v_i-\mathfrak{L}$,
is a conserved quantity that can be interpreted as the
1PN generalization of the classical energy:
\be\label{ec_ene}
E = \frac{v^2}{2} + \phi + \frac{1}{c^2}\left(\frac{3v^{4}}{8}-\frac{3v^{2}\phi}{2}+\frac{\phi^2}{2}+\psi\right).
\ee
Note that this expression is independent of the vector field $\mbox{\boldmath$\xi$}$.

If the source of gravitation is endowed with axial symmetry, the $z$-component of the angular momentum
is an additional integral of motion. In cylindrical coordinates
$(R,\varphi,z)$ we obtain that, for $\varphi$-independent potentials, the quantity
\begin{eqnarray}\label{lz_pn}
L_z = Rv_{\varphi}+\frac{1}{c^2}\left[Rv_{\varphi}\left(\frac{v^{2}}{2}-3\phi\right)+R\xi_{\varphi}\right]
\end{eqnarray}
can be interpreted as the 1PN generalization of the azimuthal angular momentum. In this case $\mbox{\boldmath$\xi$}$ plays a role,
through its rotational component.\\

\subsection{1PN statistical mechanics}

From a statistical point of view, the state of the system can be determined by its DF, $F({\bf x},{\bf v},t)$, depending on
the spatial coordinates, velocity and time. Now, as mentioned in the Introduction, for applications in galactic dynamics
it is commonly assumed that the encounters between particles are negligible and hence, the evolution of the stellar
system must obey the so-called collisionless
Boltzmann equation. In 1PN approximation, such relation can be written as \cite{pedraza-agon-ramos}
\begin{equation}\label{ec_general derezania}
\begin{split}
\qquad\qquad\frac{\partial F}{\partial t}+v^{i}\frac{\partial F}{\partial x^{i}}-
\frac{\partial \phi}{\partial x^{i}}\frac{\partial F}{\partial v^{i}}+
\frac{1}{c^2}\left(\frac{v^{2}}{2}-\phi \right)
\left(\frac{\partial F}{\partial t}+v^{i}\frac{\partial F}{\partial x^{i}}\right)\qquad\qquad\qquad\qquad\\\\
+\frac{1}{c^{2}}\left[4v^{i} v^{j}\frac{\partial \phi}{\partial x^{j}}-
\left(\frac{3v^{2}}{2}+3\phi\right)\frac{\partial \phi}{\partial x^{i}}
-v^{j}\left(\frac{\partial \xi_{i}}{\partial x^{j}}-\frac{\partial \xi_{j}}{\partial
  x^{i}}\right)+
3v^{i}\frac{\partial \phi}{\partial t}-\frac{\partial \psi}{\partial x^{i}}-\frac{\partial \xi_{i}}{\partial t}
\right]\frac{\partial F}{\partial v^i}=0.
\end{split}
\end{equation}
For situations in which encounters play a dominant role, the right-hand side of the above expression can be replaced by a collisional term of the Fokker-Planck type \cite{ramos-gonzalez}.

If the self-gravitating system is in a stationary state, $\partial F/\partial t=0$ and the DF can be expressed as a function of the energy (a first integral of motion). Moreover, if the system
is endowed with additional symmetries we can expect that, as in Newtonian theory, the stationary solutions of
(\ref{ec_general derezania}) are also functions of the remaining integrals of motion. In other words, there is a 1PN version of the Jeans theorem, which simply reflects the fact that equation (\ref{ec_general derezania}) can be rewritten as $dF/dt=0$ \cite{pedraza-agon-ramos}.

The DF describes completely the sate of the system. We
can extract from it all the relevant physical quantities, like the mass density, the mean velocity, the velocity-dispersion tensor, etc.
For instance, by taking the moments of equation (\ref{ec_general derezania}), we could in principle obtain some useful results regarding the hydrodynamics of self-gravitating systems at 1PN order. This was first derived by Chandrasekhar \cite{chan} using a different approach so we will refrain from writing out
these results here, since they are not particularly illuminating.

Another important fact that can be derived with the help of the DF, is the
1PN virial theorem (see appendix \ref{sec:ap-virial} for details). This theorem shows that there exists a linear relation between the variation
of the moment of inertia tensor $I^{ik}$, the kinetic energy tensor $K^{ik}$ and the potential energy tensor $W^{ik}$, according to:
\bea
\frac{d^2 I^{ik} }{dt^2}=2K^{ik}
+W^{ik}+ \frac12\frac{d}{d t}\int d^3x \frac{\stackrel{0}{\rho}}{c^2}\frac{\partial \phi }{\partial t}x^ix^k\,.
\eea
This 1PN virial equation has a similar form as the Newtonian one, except for the last term in the right-hand side.
The temporal variation of $\phi$ contributes to the variation of $I^{ik}$.
In particular, for stationary states we have
$$
W=-2K,
$$
as in the Newtonian case. This is, the total gravitational potential energy (a negative quantity), is two times the total kinetic energy.

\section{The 1PN Self-gravitation Equations}\label{sec:1PN-self-grav-eq}

The integration of the DF over the velocity space, leads to the energy-momentum tensor. It in turn determines the gravitational field via the Einstein equations, so it should be possible to derive a relation between $F$ and the potentials $\phi$,
$\psi$, and $\mbox{\boldmath$\xi$}$. Such relation is represented by a set of couple equations which we
shall call the \emph{1PN self-gravitation equations}. At first we  deal with the general case of
arbitrary self-gravitating systems and then we shall consider the special case of razor thin disks.

\subsection{Fundamental equations for arbitrary systems}

We begin by considering the following expression for the energy-momentum tensor, valid for any self-gravitating system:
\begin{equation}\label{tmunu}
T^{\mu\nu}(x^i,t)=\frac{1}{c}\int\frac{ U^\mu U^\nu}{U^0}F(x^i,U^i,t)\sqrt{-g}d^3U.
\end{equation}
Here $U^\nu$ is the four-velocity, which is related to the three-velocity by $U^i=U^0 v^{i}/c$ (Greek indices run from 0 to 3 and Latin indices from 1 to 3). From the
relation $g_{\mu\nu}U^\mu U^\nu=-c^2$ we can compute the quantities we need to derive the 1PN approximation
of the energy-momentum tensor.
After some calculations, we obtain
\begin{equation}\label{Ucero}
\frac{U^0}{c}=1+\frac{v^2}{2c^2}-\frac{\phi}{c^2},
\end{equation}
and
\be
d^3U=\left(1+\frac{5v^2}{2c^2}-\frac{3\phi}{c^2}\right)d^3v
\ee
Now, taking into account that $\sqrt{-g}=1-2\phi/c^2+...$, we find that
\bea\label{T1PN-1}
    T^{00}&=&\int \gamma Fd^{3}v,\nonumber\\
    T^{ij}&=&\frac{1}{c^2}\int \gamma\, v^{i}v^{j}Fd^{3}v,\\
    T^{0i}&=&\frac{1}{c}\int \gamma\, v^{i} Fd^{3}v,\nonumber
\eea
where
\begin{equation}\label{T1PN-2}
  \gamma=1+\frac{3v^{2}}{c^{2}}-\frac{6\phi}{c^{2}}
\end{equation}
is the measure of the integration over velocities.

According to the above relations, we expect that the DF can be expanded in power series of $\bar{v}/c$ as
\begin{equation}\label{DFsplit}
    F=\;\stackrel{0}{F}+\;\stackrel{2}{F}+\cdots,
\end{equation}
where $\stackrel{0}{F}$ is the Newtonian contribution to the DF (which is itself a solution of the classical
CBE) and
$\stackrel{2}{F}$ is the first post-Newtonian correction.
Plugging (\ref{DFsplit}) into (\ref{T1PN-1}) leads to the different components
of the energy-momentum tensor at the orders required by the 1PN approximation:
\begin{eqnarray}
\stackrel{0 \: \: \: \: }{T^{00}} &=&\int\,\! \stackrel{0}{F} d^{3}v, \label{T000} \\
\stackrel{2 \: \: \: \: }{T^{00}}&=& \frac{3}{c^2}\int (v^{2}
-2\phi) \;\stackrel{0}{F}d^{3}v+\int \,\!\stackrel{2}{F} d^{3}v, \label{T002}  \\
\stackrel{2 \: \: \: \: }{T^{ij}} &=& \frac{1}{c^2}\int v^{i} v^{j}\;\stackrel{0}{F}d^{3}v, \label{Tij2}  \\
\stackrel{1 \: \: \: \: }{T^{0i}} &=& \frac{1}{c}\int v^{i} \;\stackrel{0}{F}d^{3}v, \label{T0j1}
\end{eqnarray}
along with $\stackrel{0 \: \: \: \: }{T^{ij}}=0$, as expected. With these results we are ready to write the
1PN self-gravitation equations for general stationary systems:
\begin{eqnarray}
\nabla^2 \phi &=&4\pi G\int \,\!\stackrel{0}{F} d^{3}v, \label{selfgrav-1}  \\
\nabla^2 \psi &=& 8\pi G\int(2v^{2}-3\phi)\;\stackrel{0}{F}d^{3}v\nonumber\\
&&+\,4\pi G c^{2}\int\,\!\stackrel{2}{F}d^{3}v, \label{selfgrav-2}\\
\nabla^2 \xi_{i} &=& 16\pi G\int v^{i}\;\stackrel{0}{F}d^{3}v.\label{selfgrav-3}
\end{eqnarray}
To summarize, we can say that a stellar system characterized by an equilibrium
DF satisfying (\ref{ec_general derezania}) is described by
a matter distribution given by (\ref{T000})-(\ref{T0j1}), and gravitational interactions determined by the field equations
 (\ref{ec_campo_pn1})-(\ref{ec_campo_pn3}). In order to provide a self-consistent description, the relations (\ref{selfgrav-1})-(\ref{selfgrav-3})
must be satisfied. All of these equations are written as power expansions in the small parameter $\bar{v}/c$ and, in consequence,
we can clearly distinguish between the Newtonian contribution and the post-Newtonian corrections.

\subsection{The case of stationary razor thin disks}\label{sec: razor thin disk}

The so-called razor thin disks are of special interest in modeling a number of axisymmetric galaxies.
In this case, the DF depends on velocities and positions in the form
\begin{equation}\label{DF-razor}
    F=f(R,v_{R},v_{\varphi})\delta(z)\delta(v_{z}),
\end{equation}
where $\delta$ is the Dirac delta function and $f$ is a reduced phase-space density describing the stellar population
placed on the equatorial plane $z=0$ (for a finite thin disk of radius $a$, $f$ must vanish for $R>a$).
Equations (\ref{selfgrav-1}) and (\ref{selfgrav-2}) can be written as
\begin{eqnarray}
\nabla^2 \phi &=&4\pi G\delta(z)\int
 \!\,\stackrel{0}{f} d^{2}v, \label{selfgrav-1-disk} \\ &&\nonumber\\
\nabla^2 \psi &=& 4\pi G\delta(z)\left[\int (4v^{2}-6\phi)\,\stackrel{0}{f} d^{2}v
+c^{2}\int\!\,\stackrel{2}{f}d^{2}v\right]\nonumber\\\label{selfgrav-2-disk}
\end{eqnarray}
where $d^{2}v=dv_{R}dv_{\varphi}$. Equation (\ref{selfgrav-3}) requires a little more attention.
First start with equation (\ref{ec_campo_pn3}) (which is equivalent to (\ref{selfgrav-3})). The general solution that
vanishes at infinity can be written as \cite{WB}
\begin{equation}\label{soluciongenXi}
    \xi_{i}(\mathbf{x})=-4Gc\int\frac{\stackrel{1 \: \: \: \: }{T^{0i}}(\mathbf{x'})d^{3}x'}{|\mathbf{x}-\mathbf{x'}|},
\end{equation}
which, in our case, reduces to
$$
-4G\int\int \frac{v^{i}\stackrel{0}{F} d^{3}x'd^{3}v}{|\mathbf{x}-\mathbf{x'}|}
    =-4G\int\int \frac{v^{i}\stackrel{0}{f} \delta(z')d^{3}x'd^{2}v}{|\mathbf{x}-\mathbf{x'}|},
$$
where we have computed the integral with respect to $v_{z}$ (we use the notation $v^{1}=v_{x}$, $v^{2}=v_{y}$ and $v^{3}=v_{z}$).
This expression can be massaged into a useful form by taking into account two facts:
(i) the relation between the cartesian components $v_{x}$, $v_{y}$
and the cylindrical components $v_{R}$, $v_{\varphi}$, i.e. $v_{x}=v_{R}\cos\varphi-v_{\varphi}\sin\varphi$ and $v_{y}=v_{R}\sin\varphi+v_{\varphi}\cos\varphi$; (ii) since we are dealing with stationary axisymmetric systems,
the DF is an even function of $v_{R}$ \cite{BT} and in consequence $\int v_{R} fdv_{R}=0$ . Thus, we can write
$$
\int\int \frac{v_{x}\stackrel{0}{f} \delta(z')d^{3}x'd^{2}v}{|\mathbf{x}-\mathbf{x'}|}=
-\sin\varphi\int\int \frac{v_{\varphi}\stackrel{0}{f} \delta(z')d^{3}x'd^{2}v}{|\mathbf{x}-\mathbf{x'}|}
$$
and
$$
\int\int \frac{v_{y}\stackrel{0}{f} \delta(z')d^{3}x'd^{2}v}{|\mathbf{x}-\mathbf{x'}|}=
\cos\varphi\int\int \frac{v_{\varphi}\stackrel{0}{f} \delta(z')d^{3}x'd^{2}v}{|\mathbf{x}-\mathbf{x'}|}.
$$
Now, by introducing the relations $\xi_{x}=\xi_{R}\cos\varphi-\xi_{\varphi}\sin\varphi$ and $\xi_{y}=\xi_{R}\sin\varphi+\xi_{\varphi}\cos\varphi$
in (\ref{soluciongenXi}), we obtain
$$
\xi_{R}\cos\varphi-\xi_{\varphi}\sin\varphi=\sin\varphi\int\int 4G\frac{v_{\varphi}\stackrel{0}{f} \delta(z')d^{3}x'd^{2}v}{|\mathbf{x}-\mathbf{x'}|}
$$
and
$$
\xi_{R}\sin\varphi+\xi_{\varphi}\cos\varphi=-\cos\varphi\int\int 4G\frac{v_{\varphi}\stackrel{0}{f} \delta(z')d^{3}x'd^{2}v}{|\mathbf{x}-\mathbf{x'}|}.
$$
Since we assume that $\mbox{\boldmath$\xi$}$ is $\varphi$-independent, each of these expressions leads us to the conclusion that
\begin{equation}\label{xiR-finite-dist}
    \xi_{R}=0,\qquad\mbox{for finite distributions,}
\end{equation}
and that $\xi_{\varphi}$ is solution of the following equation:
\begin{equation}
\nabla^2 \xi_{\varphi} =16\pi G\delta(z)\int
 v_{\varphi}\!\,\stackrel{0}{f} d^{2}v. \label{selfgrav-3a-disk}
\end{equation}
The equation for the component $\xi_{z}$ can be obtained easily by replacing (\ref{DF-razor}) in
(\ref{selfgrav-3}), and the result is the Laplace equation,
$\nabla^{2}\xi_{z}=0.$ Its solution can be determined through condition (\ref{condition-phi-xi}). A straightforward
calculation leads to
\begin{equation}\label{xiz}
    \xi_{z}(R)=\xi_{zo}\ln(R/R_{o}),
\end{equation}
where $\xi_{zo}$ and $R_{o}$ are constants of integration. In the case of distributions with finite extent, we demand as a boundary condition that
$\lim_{R\rightarrow\infty}\xi_{z}=0$. In consequence, we have to choose $\xi_{zo}=0$, and hence
\begin{equation}\label{xiz-finite-disks}
    \xi_{z}=0,\qquad\mbox{for finite thin disks.}
\end{equation}

On the other hand, we expect that $f$ obeys the collisionless  Boltzmann equation in
the three-dimensional phase-space $(R,v_{R},v_{\varphi})$. In fact,  by introducing (\ref{DF-razor})
in (\ref{ec_general derezania})
and performing an integration on $z$ and $v_{z}$,
it follows that the distribution $f$ obeys the relation
\bea\label{ec_derezania cilindricas}
-\left(1+\frac{v^{2}}{2c^{2}}-\frac{\phi}{c^{2}}-
\frac{4R}{c^{2}}\frac{\partial \phi}{\partial R}
-\frac{R}{c^{2}v_{\varphi}}\frac{\partial\xi_{\varphi}}{\partial R}\right)
\frac{v_{R}v_{\varphi}}{R}\frac{\partial f}{\partial v_{\varphi}}+\left(1+\frac{v^{2}}{2c^{2}}-
\frac{\phi}{c^{2}}\right)v_{R} \frac{\partial f}{\partial R}&&
\nonumber\\ \\
+\left[\left(1+\frac{v^{2}}{2c^{2}}-\frac{\phi}{c^{2}}\right)\frac{v_{\varphi}^{2}}{R}
\left(1+\frac{3v_{\varphi}^{2}-5v_{R}^{2}}{2c^{2}}+\frac{3\phi}{c^{2}}\right)\frac{\partial \phi}{\partial R}-
\frac{1}{c^{2}}\frac{\partial \psi}{\partial R}
+\frac{v_{\varphi}}{c^{2}}\frac{\partial\xi_{\varphi}}{\partial R}\right]\frac{\partial f}{\partial v_{R}}&=&0,\nonumber
\eea
where $v^{2}=v^{2}_{R}+v^{2}_{\varphi}$ and $\partial\phi/\partial R$, $\partial\psi/\partial R$ and
$\partial\xi_{\varphi}/\partial R$ are evaluated at $z=0$.
The above relation is the 1PN version of the Boltzmann equation for an axisymmetric
two-dimensional shell, located at the equatorial plane, and in a stationary state. Of course, $f(R,v_{R},v_{\varphi})$
plays the role of the reduced DF describing the diskoidal shell.

It is straightforward to show that $E$ and $L_z$, given by equations (\ref{ec_ene}) and (\ref{lz_pn}), are solutions
of (\ref{ec_derezania cilindricas}).
 This means that, for axially symmetric systems, any $f$ depending on $E$ and $L_{z}$ is
solution of the CBE; conversely, any solution of
the CBE can always be expressed as a function of
$E$ and $L_{z}$. Thus, any two-integral DF, $f(E,L_z)$, provides a complete statistical description
for the (two-degree-of-freedom) stellar system. This fact will be very useful for the formulation of
post-Newtonian models in the next section.

\section{Analytical Models for Axisymmetric Galaxies}\label{sec:analyitical-models}

The purpose of this section is to show how to implement the formalism developed above in order to obtain axially symmetric galaxy models.
For the applications we want to consider here we have to take into account further considerations.
First of all, recall that in \cite{pedraza-agon-ramos} we proved that Jeans theorem remains valid at 1PN order.
This means that any equilibrium solution of the CBE depends only on the integrals of motion of the system, and that any function
of the integrals yields an equilibrium solution of the CBE. Thus, for stationary systems with axial symmetry,
we can restrict ourselves to DFs depending on the energy (\ref{ec_ene}) and the angular momentum (\ref{lz_pn}),
which are themselves integrals of (\ref{ec_general derezania}).

The next step would be to implement the previous restrictions starting from a given Newtonian potential-density pair with a known DF,
as was done in \cite{pedraza-agon-ramos} for the spherically symmetric case. As a result, one expects two coupled
\emph{self-gravitation equations}, providing a method to determine, from a Newtonian model, its associated post-Newtonian corrections.
In practice, the present formalism leads to a two coupled ordinary differential equations in the spherically symmetric case. In the axially symmetric situation however, one ends up with two coupled elliptic partial
differential equations (for general volumetric matter distributions).

In this case such equations are much
more involved than the ones corresponding to the spherically symmetric case,
but the configurations we shall deal here permit us to introduce some additional assumptions to simplify the problem.
In the next section, we shall show that a dramatic simplification
can be achieved by the consideration of thin discoidal distributions in spheroidal oblate coordinates:
instead of getting differential equations, in this case the post-Newtonian corrections can be obtained from simple algebraic
equations.

We then present a particular application where the resulting
equations can be solved analytically, which means that it is possible to obtain \emph{1PN exact solutions}.
The importance of these solutions will be evaluated by a comparison between
density profiles and rotation curves described by Newtonian theory and the ones predicted by the 1PN
approximation. Although focus on the particular models introduced in
\cite{MYM} (revisited by \cite{GR}), our framework can be applied to a wider variety of models.
In general, the method can be used for situations in which the potentials are separable
functions of the spheroidal oblate coordinates.

\subsection{Hunter's method in the 1PN approximation} \label{hunterms meth 1PN}

One can find in the literature a number of self-consistent stellar models representing razor thin disks; here we will deal with
models belonging to the family of Morgan \& Morgan disks \cite{MYM,GR}. In the
Newtonian formulation, they can be obtained by a formalism developed by Hunter \cite{HUN1}.
Such procedure (known as the Hunter's method)
provides the surface density of the disks,
the gravitational potential, and the circular velocity as a series of elementary
functions, by superposing solutions of the Laplace
equation in oblate spheroidal coordinates.
Hunter's method can also be implemented in the context of
the 1PN approximation as follows.

To begin with, note that in vacuum, the field equations
(\ref{ec_campo_pn1})-(\ref{ec_campo_pn3}) reduce to three
Laplace equations for  $\phi$, $\psi$, and $\xi_{\varphi}$ (remember that $\xi_{z}=0$ for distributions
of finite extent). Without loss of generality, we assume that the disk is on the equatorial plane, so
we have to impose that the gravitational potentials have symmetry of reflection with respect
to the plane $z=0$, i.e., $\phi(R,z)= \phi(R,-z)$, $\psi(R,z)= \psi(R,-z)$, and $\xi_{\varphi}(R,z)= \xi_{\varphi}(R,-z)$. Then, it follows that
\begin{eqnarray}
&& \frac{\partial\phi}{\partial z}(R,-z)= -\frac{\partial\phi}{\partial z}(R,z), \label{reflection2a}\\
&&\nonumber\\
&& \frac{\partial\psi}{\partial z}(R,-z)= -\frac{\partial\psi}{\partial z}(R,z), \label{reflection2b}\\
&&\nonumber\\
&& \frac{\partial\xi_{\varphi}}{\partial z}(R,-z)= -\frac{\partial\xi_{\varphi}}{\partial z}(R,z), \label{reflection2c}
\end{eqnarray}
in agreement with the attractive character of  gravitation. We also assume that $\partial\phi/\partial z$,
$\partial\psi/\partial z$, and $\partial\xi_{\varphi}/\partial z$ do not vanish in the disk's zone, in order to have the corresponding
thin distribution of energy-momentum. Such distribution, restricted to a region $0\leq R \leq a$ in the plane $z=0$
(from here on, $a$ will denote the disk radius), will be described by a ``shell-like" energy-momentum
tensor. If we define
\begin{eqnarray}
\stackrel{0 \: \: \: \: }{T^{00}} &=& {\Sigma}(R) \delta(z),\\
\stackrel{2 \: \: \: \: }{T^{00}} + \stackrel{2 \: \: \: \: }{T^{ii}} &=& \frac{1}{c^2}{\sigma}(R)
\delta(z),\\
\stackrel{1 \: \: \: \: }{T^{0\varphi}} &=& \frac{1}{c}{\Delta}(R) \delta(z),
\end{eqnarray}
for $0\leq R \leq a$, it follows from Gauss's Law that
\begin{eqnarray}
&& {\Sigma}(R) = \frac{1}{2 \pi G}\left(\frac{\partial \phi}{\partial z}\right)_{z=0^{+}}, \label{Sigma-phi}\\
&& \nonumber\\
&& {\sigma}(R) = \frac{1}{2 \pi G}\left(\frac{\partial \psi}{\partial z}\right)_{z=0^{+}}, \label{sigma-psi}\\
&& \nonumber\\
&& {\Delta}(R) = \frac{1}{8 \pi G}\left(\frac{\partial \xi_{\varphi}}{\partial z}\right)_{z=0^{+}}. \label{sigma-xi}
\end{eqnarray}
Note that ${\Sigma}$ represents the surface mass density of the Newtonian theory (i.e. without relativistic corrections),
 while ${\Delta}$ plays the role of the surface density
of $\varphi$-momentum. On the other hand, ${\sigma}$ is associated both  to the pressure and the relativistic corrections to the
mass surface density.

The above relations mean that, in order to have a distribution
of matter as the described by (\ref{Sigma-phi})-(\ref{sigma-xi}), we have to demand that
\begin{eqnarray}
&&\frac{\partial \phi}{\partial z}(R,0^{+}) \neq 0 , \qquad R\leq a, \label{con1} \\
&& \nonumber\\
&&\frac{\partial \phi}{\partial z}(R,0^{+}) = 0 , \qquad R > a,\label{con2}
\end{eqnarray}
with the same requirement for $\psi$ and $\xi_{\varphi}$. At this point it is convenient to introduce  oblate
spheroidal coordinates, a system that adapts in a
natural way to the geometry of the problem. They are related to the cylindrical ones through
\begin{eqnarray}
&& R = a \sqrt{(1+\zeta^{2}) (1-\eta^{2})}, \\
&& z = a \zeta \eta,
\end{eqnarray}
where $0 \leq \zeta<\infty$ and $-1\leq \eta< 1$. Note that (i) the disk itself has coordinates
$\zeta = 0$, $\eta^{2}=1-R^{2}/a^{2}$; (ii) conditions (\ref{con1})-(\ref{con2}) become
\begin{eqnarray}
&&\frac{\partial \phi}{\partial \zeta}\bigg|_{\zeta=0}=H(\eta), \label{con3} \\
&& \nonumber\\
&&\frac{\partial \phi}{\partial \eta}\bigg|_{\eta=0} = 0,\label{con4}
\end{eqnarray}
where $H$ is an even function of $\eta$. The general solution of Laplace's equation satisfying the above conditions
can be written as
\begin{equation}
\phi(\zeta,\eta) = - \sum_{n=0}^{\infty} A_{2n} q_{2n}(\zeta) P_{2n}(\eta),
\label{phi1}
\end{equation}
where $A_{2n}$ are arbitrary constants, $P_{2n}(\eta)$ and $q_{2n}(\zeta)=
i^{2n+1}Q_{2n}(i\zeta)$ are the usual Legendre polynomials and the Legendre
functions of second kind, respectively. The post-Newtonian potentials $\psi$
and $\xi_{\varphi}$ have the same form,
\begin{equation}
\psi(\zeta,\eta) = - \sum_{n=0}^{\infty} B_{2n} q_{2n}(\zeta) P_{2n}(\eta),
\label{psi1}
\end{equation}
\begin{equation}
\xi_{\varphi}(\zeta,\eta) = - \sum_{n=0}^{\infty} C_{2n} q_{2n}(\zeta) P_{2n}(\eta),
\label{xi1}
\end{equation}
but here we have denoted the expansion constants as $B_{2n}$ and $C_{2n}$.
We can derive explicit formulas
for $\Sigma$, $\sigma$, and $\Delta$ in oblate spheroidal coordinates,
by introducing (\ref{phi1})-(\ref{xi1}) in (\ref{Sigma-phi})-(\ref{sigma-xi}):
\begin{eqnarray}
\Sigma = \frac{1}{2\pi a G \eta_{*}}\sum_{n = 0}^{\infty} A_{2n} (2n+1)
q_{2n+1}(0) P_{2n}(\eta_{*}),&& \label{Sigma1}\\&&\nonumber\\
\sigma = \frac{1}{2\pi a G \eta_{*}}\sum_{n = 0}^{\infty} B_{2n} (2n+1)
q_{2n+1}(0) P_{2n}(\eta_{*}),&&  \label{sigma1}\\&&\nonumber\\
\Delta = \frac{1}{8\pi a G \eta_{*}}\sum_{n = 0}^{\infty} C_{2n} (2n+1)
q_{2n+1}(0) P_{2n}(\eta_{*}),&&
\end{eqnarray}
where $\eta_{*}$ represents the value of coordinate $\eta$ inside the disk:
\begin{equation}\label{eta0}
    \eta_{*}=\sqrt{1-\frac{R^{2}}{a^{2}}}
\end{equation}

Now that we have stated the fundamental structure
of models with 1PN corrections, the next step is to demand that the models obtained
are self-consistent, i.e., that they have an analytical equilibrium DF that
is related consistently to the surface mass distribution. In other words, we have to formulate
the corresponding 1PN self-gravitating equations:
\begin{eqnarray}
\sum_{n = 0}^{\infty} \tilde{A}_{2n} \frac{P_{2n}(\eta_{*})}{\eta_{*}} &=&\int
\,\!\stackrel{0}{f} d^{2}v, \label{selfgrav-1-morgan} \\ &&\nonumber\\
\sum_{n = 0}^{\infty} \tilde{B}_{2n} \frac{P_{2n}(\eta_{*})}{\eta_{*}} &=&\int (4v^{2}-6\phi)\,\stackrel{0}{f} d^{2}v
+c^{2}\int\,\!\stackrel{2}{f}d^{2}v,\nonumber\label{selfgrav-2-morgan}\\&&\\
\sum_{n = 0}^{\infty} \tilde{C}_{2n} \frac{P_{2n}(\eta_{*})}{\eta_{*}} &=&\int
 v_{\varphi}\,\!\stackrel{0}{f} d^{2}v, \label{selfgrav-3-morgan}
\end{eqnarray}
where, for the sake of simplicity, we have defined
\begin{equation}\label{constantes A}
    \tilde{A}_{2n}=\frac{(2n+1)q_{2n+1}(0)}{2\pi a G}A_{2n},
\end{equation}
the same for $\tilde{B}_{2n}$, and
\begin{equation}\label{constantes C}
    \tilde{C}_{2n}=\frac{(2n+1)q_{2n+1}(0)}{8\pi a G}C_{2n}.
\end{equation}
Note that, part of the right-hand side of equations (\ref{selfgrav-1-morgan})-(\ref{selfgrav-3-morgan})
is characterized by three fundamental quantities in Newtonian dynamics of self-gravitating
systems: (i) the mass surface density, $\Sigma=\int\,\!\stackrel{0}{f} d^{2}v$;
(ii) the mean square velocity, $\langle v^{2} \rangle=\Sigma^{-1}\int v^{2}\,\stackrel{0}{f} d^{2}v$,
and (iii) the mean circular velocity, $\langle v_{\varphi} \rangle=\Sigma^{-1}\int v_{\varphi}\,\!\stackrel{0}{f} d^{2}v$.
In particular, equation (\ref{selfgrav-1-morgan}) is the Newtonian self-gravitation equation.

There is a variety of cases that can be addressed by the formalism presented above. They correspond
to situations in which the gravitational fields (i) have symmetry of reflection with respect to the equatorial plane,
(ii) are separable functions of the spheroidal oblate coordinates and (iii) correspond to axially symmetric distributions
with finite extent. Note that any function with these features can be expressed in the form (\ref{phi1}).

\subsection{The Morgan \& Morgan disks with 1PN corrections}\label{sec:kalnajs general}

In Newtonian gravity, the Generalized Kalnajs Disks \cite{GR}
are finite thin distributions of matter with surface mass density
\begin{equation}
\Sigma_{m}(R) = \frac{(2m+1)M}{2\pi a^{2}} \left( 1 - \frac{R^{2}}{a^{2}}
\right)^{m-1/2},\qquad m\in\mathcal{N}\label{densidad-kalgen}
\end{equation}
and gravitational potential $\phi_{m}$,
given by (\ref{phi1}), with the following expansion constants
\begin{equation}
A_{2n}^{(m)}= \frac{M G}{a}\frac{\sqrt{\pi} 2^{-2m-1} (4n+1) (2m+1)!}{ (2n+1) (m - n)!
\Gamma(m + n + \frac{3}{2} )q_{2n+1}(0)}.\label{constantes-kalgen}
\end{equation}
These solutions were first obtained by Morgan \& Morgan (MM) \cite{MYM}, by solving a boundary value problem in the context
of GR. Even so, we refer to the above solutions as  MM disks for historical reasons,
but keeping in mind that we are dealing with Newtonian gravity.

All the members of this family have interesting features: a monotonically
decreasing mass density and a Keplerian rotation curve.
The case $m=1$, corresponding to the well-known Kalnajs disk \cite{KAL} (see also \cite{BT}), is an exception because it describes a self-gravitating
disk which rotates as a rigid body.

The superposition
of members belonging to the MM family, has been used to obtain new models
with more realistic properties. For example, in reference \cite{PRG}, the authors constructed a family of models with
approximately flat rotation curves considering a particular combination of MM disks.
Another example is the family of models introduced by Letelier in \cite{lete-flat-ring}, representing flat rings  (see also \cite{rpl} for astrophysical applications). Additionally, in reference \cite{gonzalez-plata-ramos} it was shown that it is possible to construct models
with realistic rotational curves obeying simple polynomial expressions. In particular, the authors constructed models for a number of galaxies of the Ursa Major cluster, and as an application they estimated their corresponding mass distributions. This fact suggest that there exist superpositions of MM members
leading to galactic models in agreement with the so-called \emph{maximum disk hypothesis}
\cite{BT}. It would be interesting to have in hand the 1PN version of the MM family, along with all of its
features.

Here we illustrate how to obtain the post-Newtonian version of the MM disks. We focus in the $m=2$ model, which is characterized by a density
\begin{equation}\label{dens-kalnajs2}
    \Sigma=\frac{5 M}{2\pi a^{2}}\eta_{*}^{3},
\end{equation}
but in principle, the same procedure can be applied to any of the remaining members (or linear combinations). From here on, we will drop the subindex of $\Sigma_{m}$, for simplicity.

It can be
shown that the above surface density can be obtained from the DF \cite{PRG2}
\begin{equation}
f(E,L_z) = k(\Omega L_z -E-5\:\Omega^{2}a^{2}/4)^{-1/4},\label{DF-kalnajs2}
\end{equation}
where
\begin{equation}\label{def-k-omega}
    k=\frac{2}{\sqrt{3}}\left[\frac{10 M}{\pi^{11}a^{5}G^{3}}\right]^{1/4},
    \qquad \Omega=\sqrt{\frac{15\pi GM}{32 a^{3}}}.
\end{equation}
Note that the DF defined by (\ref{DF-kalnajs2}) is function of the  Jacobi's integral,
$E_{J}=E-\Omega L_z$, which can be interpreted as the energy measured
from a frame that rotates with constant angular speed $\Omega$ (see Appendix \ref{ap1}).

In order to obtain a 1PN version of the model, we
start from the DF given by (\ref{DF-kalnajs2}) but using the
post-Newtonian expressions for $E$ and $L_z$, i.e., equations (\ref{ec_ene}) and (\ref{lz_pn}).
Thus, at 1PN order we can write
\begin{equation}\label{f0f2}
    f=\,\!\stackrel{0}{f}+\,\!\stackrel{2}{f},
\end{equation}
with
\be
\stackrel{0}{f}=kJ_{0}^{-1/4}
    \quad\mbox{and}\quad \,\!\stackrel{2}{f}=-\frac{k}{4}J_{2}J_{0}^{-5/4},
\ee
where
\begin{eqnarray}
 J_{0}&=&-\phi-v^{2}/2+\Omega R v_{\varphi}-5\Omega^{2}a^{2}/4,\label{J0}\\&&\nonumber\\
    J_{2}&=&-c^{-2}\left[\frac{\phi^{2}}{2}+\psi-\frac{3v^{2}\phi}{2}+\frac{3v^{4}}{8}-\Omega Rv_{\varphi}\left(\frac{v^{2}}{2}-3\phi\right)-\Omega R \xi_{\varphi}\right].\label{J2}
\end{eqnarray}
These relations determine the 1PN self-gravitation equations through (\ref{selfgrav-1-morgan})-(\ref{selfgrav-3-morgan}) and the integrals
\begin{eqnarray}
    \int \,\!\stackrel{0}{f}d^{2}v&=&\frac{5 M}{2\pi a^{2}}\eta_{*}^{3},\label{I0}\\&&\nonumber\\
    \int v^{2}\,\stackrel{0}{f}d^{2}v&=&\frac{75 G M^{2}}{448 a^{3}}\left(7-7\eta_{*}^{2}+
    6\eta_{*}^{4}\right)\eta_{*}^{3}\label{I1}\\&&\nonumber\\
    \int v_{\varphi}\,\!\stackrel{0}{f}d^{2}v&=&\frac{5}{8}
    \sqrt{\frac{15 G M^{3}}{2\pi a^{5}}}\sqrt{(1-\eta_{*}^{2})}\eta_{*}^{3}\label{Iphi}\\&&\nonumber\\
    \int \,\!\stackrel{2}{f}d^{2}v&=&\frac{1}{c^{2}\eta_{*}}\bigg[\frac{GM^{2}}{a^{3}}
    \sum_{j=0}^{4}b_{2j}\eta_{*}^{2j}+\sqrt{\frac{160 M}{3\pi^{3}a^{3}G}}\sqrt{(1-\eta_{*}^{2})}\xi_{\varphi}
    +\frac{32 \psi}{32aG\pi^{2}}\bigg]\label{I2}
\end{eqnarray}
where
\begin{eqnarray}
b_{0}=-\frac{2325}{1024}, \qquad b_{2}=\frac{75}{512},\qquad b_{4}=\frac{1125}{128},\qquad b_{6}=\frac{1275}{512},\qquad b_{8}=-\frac{38475}{7168}.\label{constantes I2}
\end{eqnarray}
By introducing (\ref{I0})-(\ref{I2}) in   (\ref{selfgrav-1-morgan})-(\ref{selfgrav-3-morgan}), we
obtain a system of linear equations for the constants $B_{2n}$ and $C_{2n}$ in terms of $A_{2n}$ (which are known a priori),
that can be solved analytically. After some computations we find that
\begin{eqnarray}
    C_{2n}&=&\left(\frac{GM}{a}\right)^{3/2}
    \frac{(45-48n-92n^{2}+16n^{3}+16n^{4})(4n+1)}{\Gamma(\frac{1}{2}-n)\Gamma(\frac{7}{2}-n)\Gamma(1+n)\Gamma(4+n)(2n+1)q_{2n+1}(0)}\frac{5 \pi^{5/2}\sqrt{15}}{256\sqrt{2}}\label{C2n}\\&&\nonumber\\
    B_{2n}&=&\left(\frac{GM}{a}\right)^{2}\frac{I_{2n}-\sum_{m=0}^{\infty} \Pi_{2n,2m}}{(64/3)q_{2n}(0)+\pi (2n+1)q_{2n+1}(0)},\label{B2n}
\end{eqnarray}
where $I_{2n}$ and $\Pi_{2n,2m}$ are constants defined in the Appendix \ref{ap2}, equations (\ref{I2n}) and (\ref{Pi2n}), respectively.

With the constants $C_{2n}$ and $B_{2n}$ at hand, the post-Newtonian fields are completely determined along with
the remaining physical quantities. In galactic dynamics, the circular velocity and the mass profile are two important
measurable quantities used to verify a particular model. In our case, the first one is given by the relation (\ref{rotlaw}) whereas the second one is given by (\ref{dens-kalnajs2}) plus the 1PN corrections coming from $\stackrel{2 \: \: \: \: }{T^{00}}$.
The corrected surface density can be written explicitly as
\be\label{dens-kalnajs1PN}
    \stackrel{0}{\Sigma}+\stackrel{2}{\Sigma} =\frac{5 M}{2\pi a^{2}}\left[\eta_{*}^{3}-
    \frac{15\pi\lambda}{224}\left(7-7\eta_{*}^{2}+64\eta_{*}^{4}\right)\eta_{*}^{3}+\frac{\lambda}{5\eta_{*}}\sum_{n=0}^{\infty}\frac{(2n)!!\hat{B}_{2n}}{(2n-1)!!}P_{2n}(\eta_{*})\right],
\ee
where $\lambda$ is a dimensionless parameter defined by
\begin{equation}\label{lambda}
    \lambda\equiv\frac{GM}{ac^{2}},
\end{equation}
 and $\hat{B}_{2n}$ are the dimensionless constants
\begin{equation}\label{otrasB}
    \hat{B}_{2n}=\frac{a^{2}B_{2n}}{G^{2}M^{2}}.
\end{equation}

\begin{figure*}
$$
\begin{array}{cc}
 \quad\quad(a) & \quad\quad(b)\\
 \epsfig{width=7cm,file=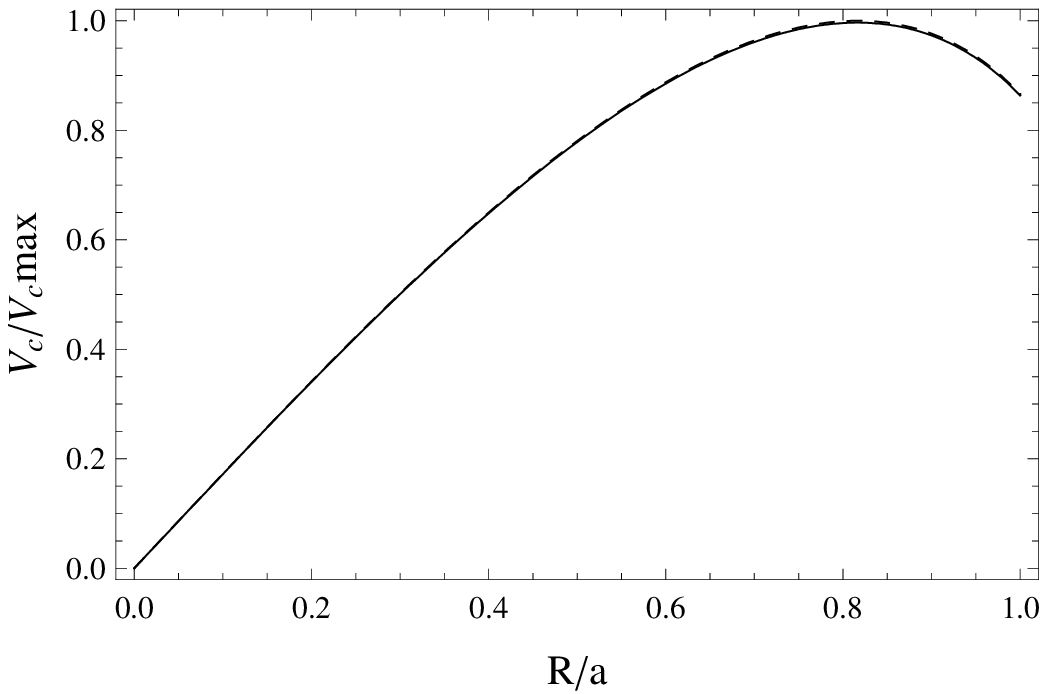} & \epsfig{width=7cm,file=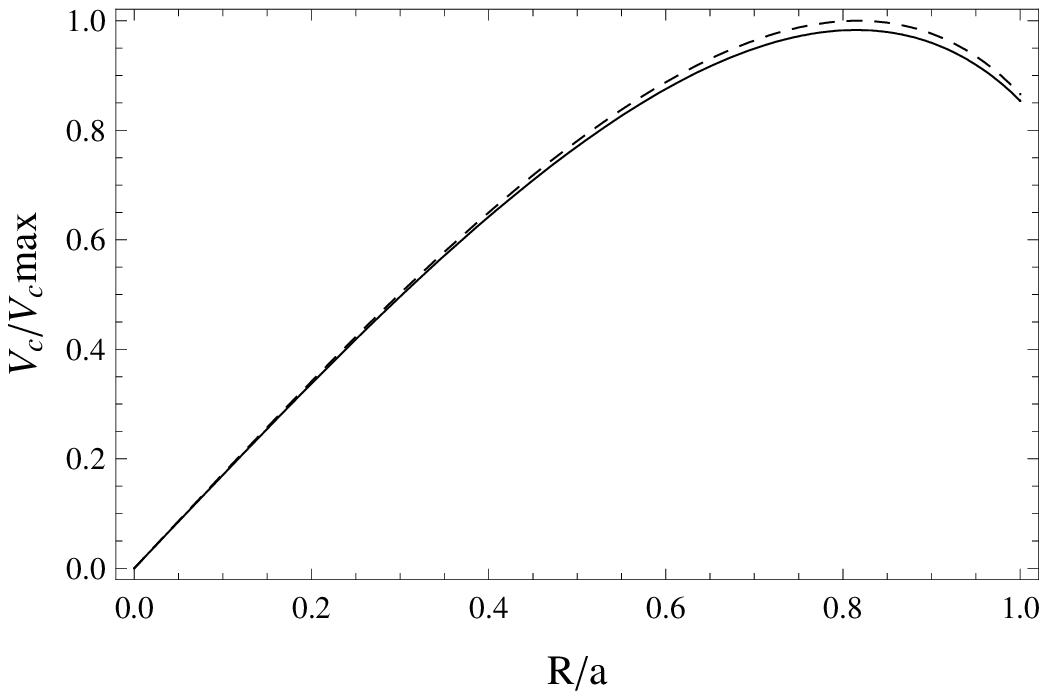}\\
 \quad\quad(c) & \quad\quad(d)\\
  \epsfig{width=7cm,file=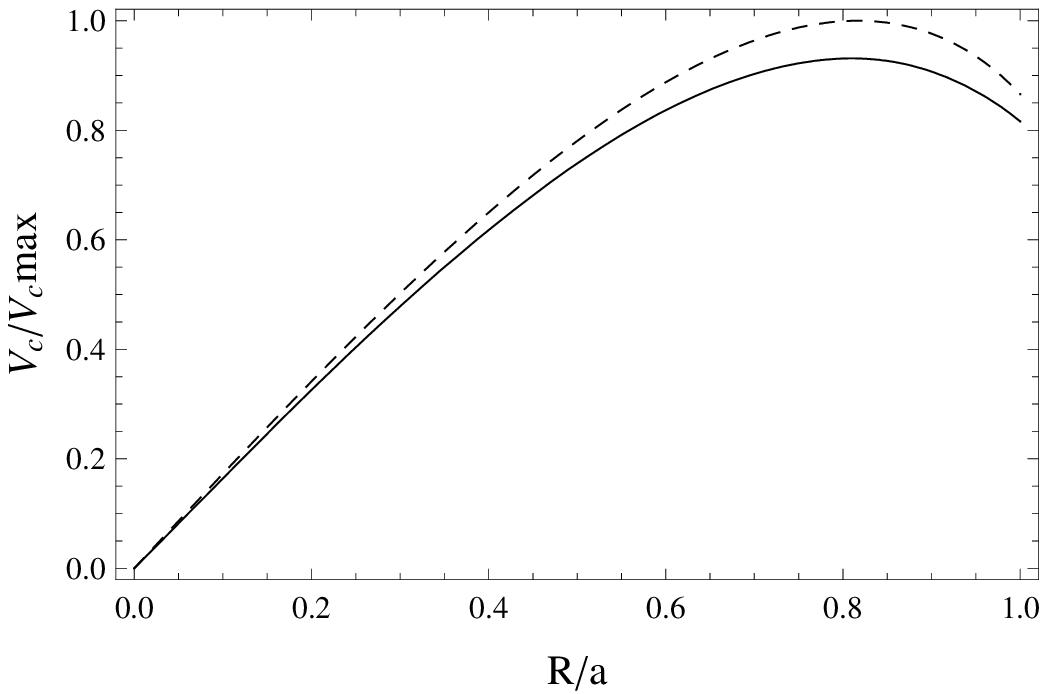} & \epsfig{width=7cm,file=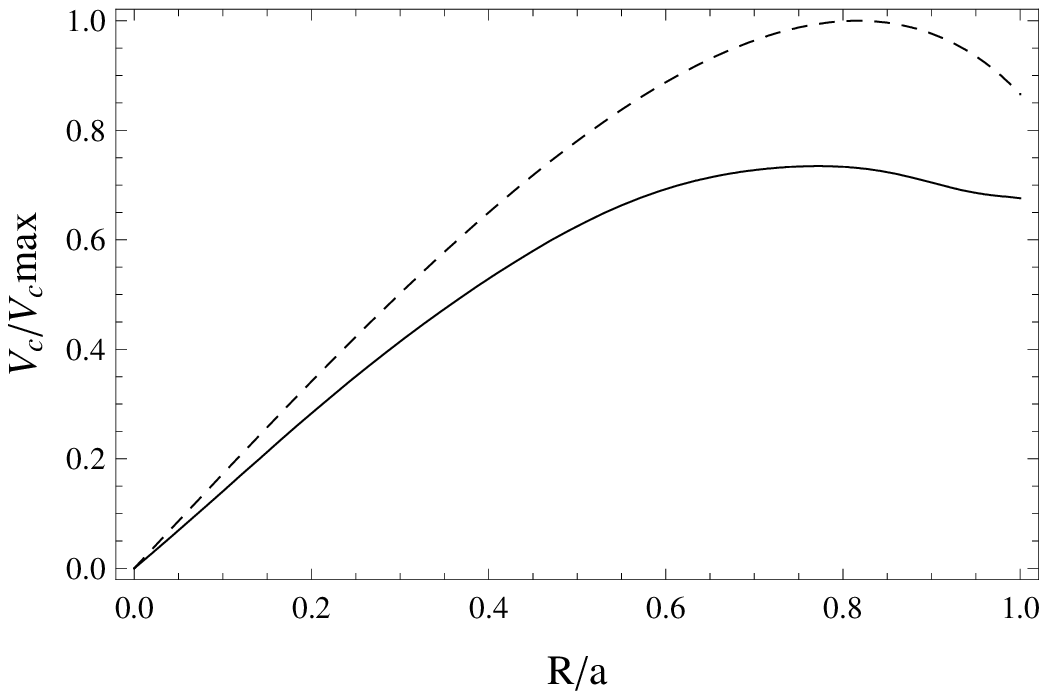}
\end{array}
$$
\caption{Rotation curves for the 1PN version of the
second MM member for different values of the
parameter $\lambda$: (a) $\lambda=10^{-3}$, (b) $\lambda=5\times10^{-3}$, (c) $\lambda=2\times10^{-2}$, and
(d) $\lambda=7\times10^{-2}$. For $\lambda\sim10^{-2}$ or more, the differences with
the Newtonian model (dashed line) become significant. In this case, the rotation curve tends to flatten as $\lambda$ grows.}
\label{fig:vel-circ-kalnajs}
\end{figure*}

\begin{figure*}
$$
\begin{array}{cc}
 \quad\quad(a) & \quad\quad(b)\\
 \epsfig{width=7cm,file=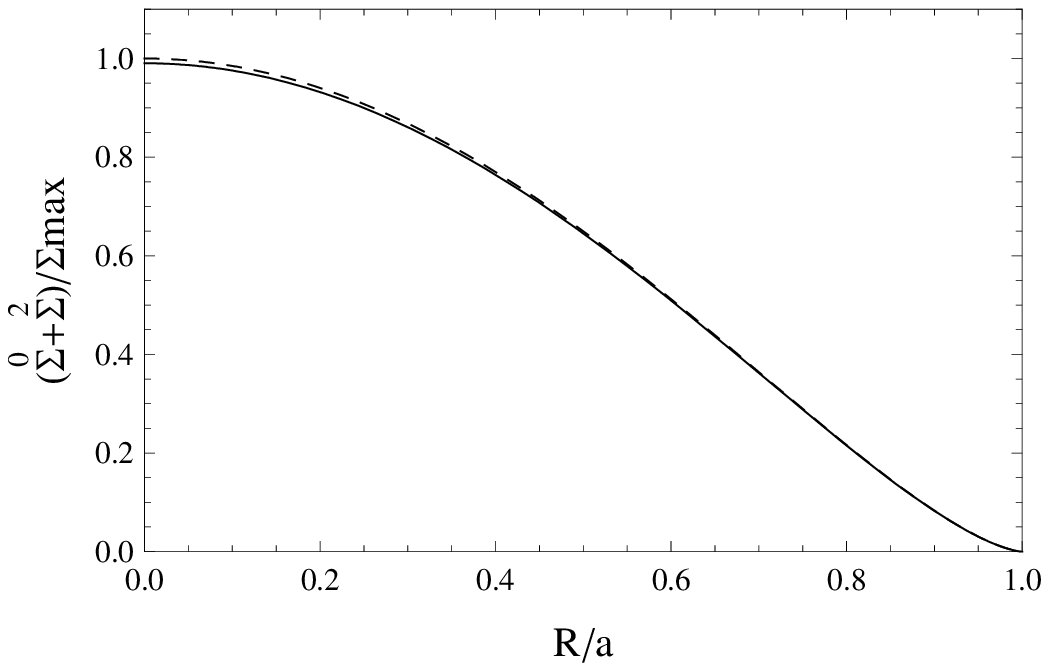} & \epsfig{width=7cm,file=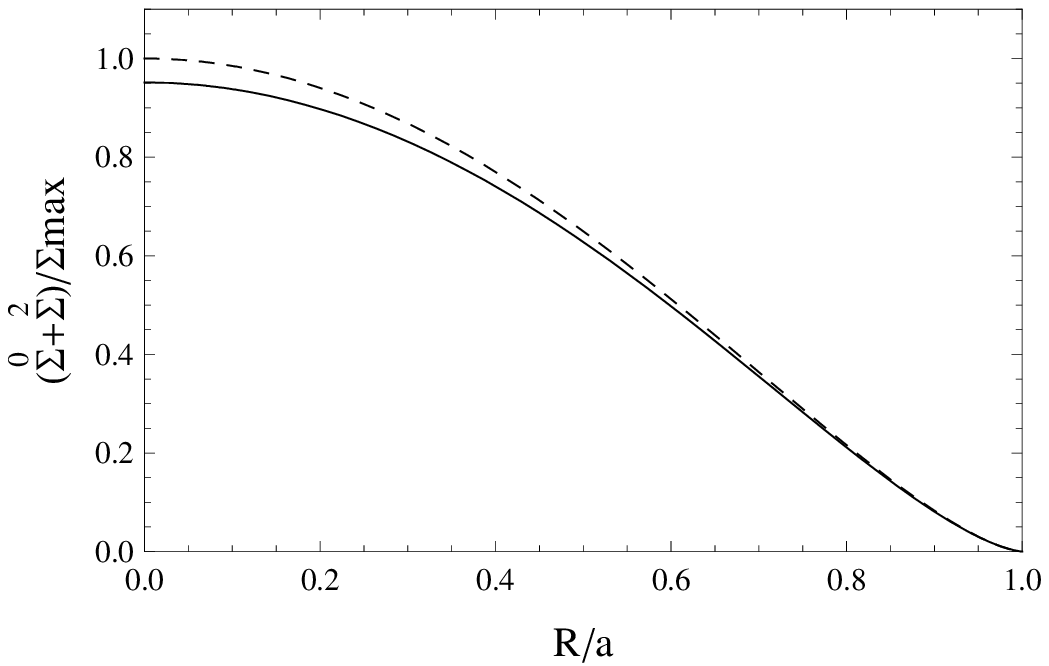}\\
 \quad\quad(c) & \quad\quad(d)\\
 \epsfig{width=7cm,file=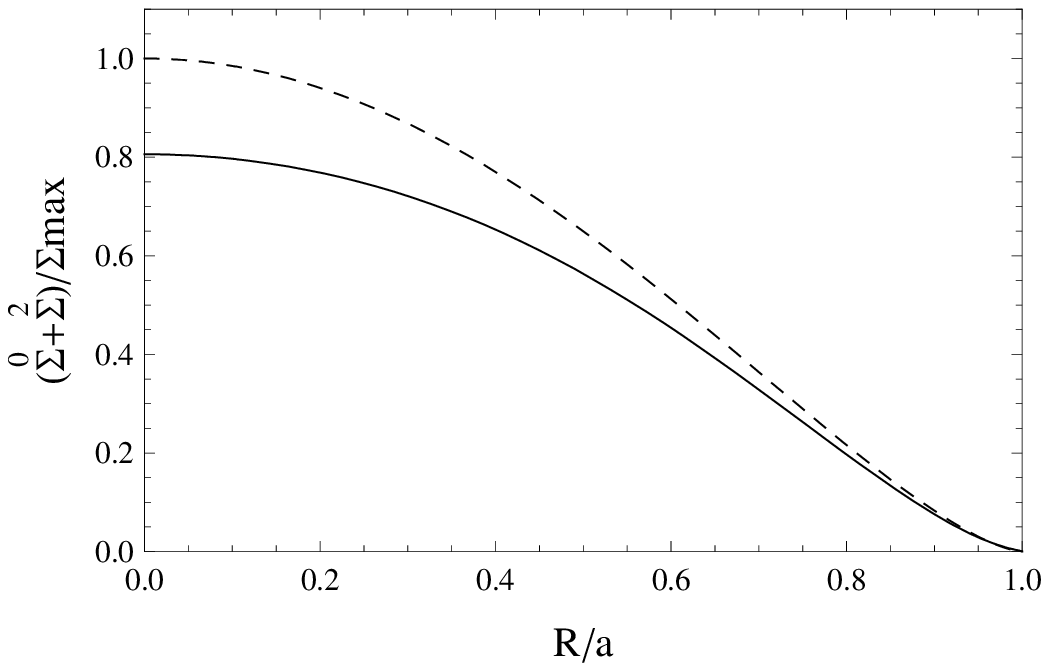} & \epsfig{width=7cm,file=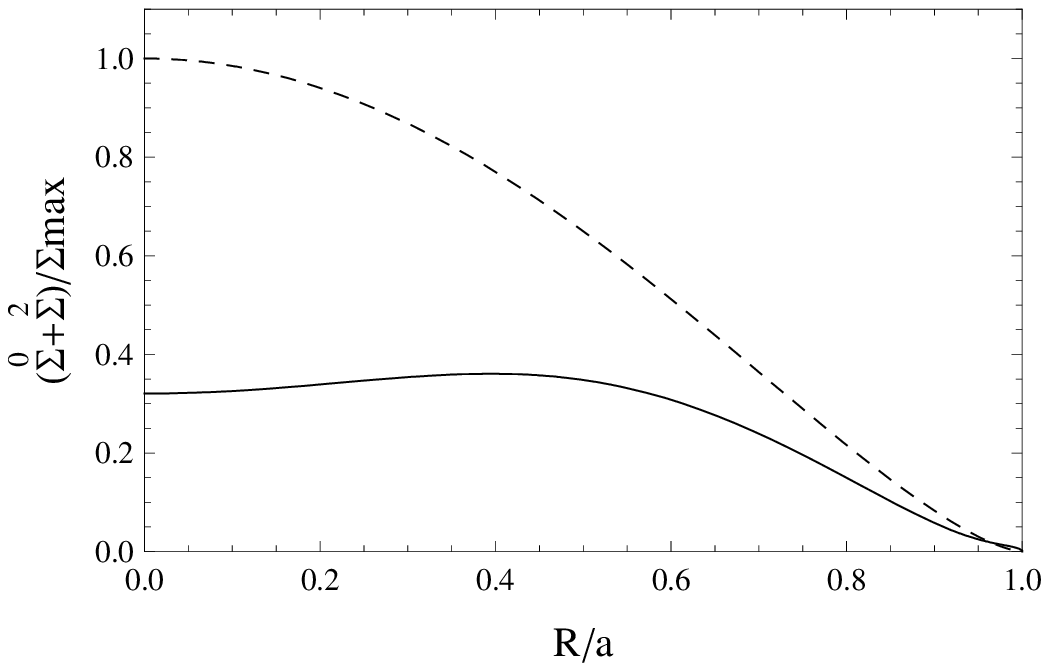}
\end{array}
$$
\caption{Mass profiles for the 1PN version of the
second MM member for the same values of the
figure \ref{fig:vel-circ-kalnajs}. As expected, the differences with
the Newtonian model (dashed line) become significant for $\lambda\sim10^{-2}$ or more. In the case (d), we have a profile in which the maximum is not at $R=0$.}
\label{fig:mass-kalnajs}
\end{figure*}

The parameter $\lambda$, which is also present in the expression for the circular velocity, is a measure
of how large are the 1PN corrections. For example, in a galaxy with $10^{12}$ solar masses and a radius
of 10 Kpc, we have  $\lambda\approx 5\times10^{-6}$. Here, we consider situations where the relativistic corrections are
larger and they can be visualized in the behavior of rotation curves and mass profile. In Figure \ref{fig:vel-circ-kalnajs},
we plot the circular velocity when $\lambda\sim10^{-3}$ and $\lambda\sim10^{-2}$. The 1PN corrections become important
in the latter case (last two figures), in particular for
values of $v_{\varphi}$ near to the disk edge. This is somewhat surprising since one would expect major
corrections near the galaxy core, where the mass concentration is maximum.
A similar phenomenon occurs with the mass profiles (see Figure \ref{fig:mass-kalnajs}). Their differences with the Newtonian profile become significant for $\lambda\sim10^{-2}$ and the magnitude of 1PN corrections increase with the radius.

\section{Concluding Remarks}\label{Conclusions}

We continued our study of the post-Newtonian kinetic theory of collisionless self-gravitating gases,
extending our previous results \cite{pedraza-agon-ramos} to systems with axial symmetry.
Before dealing with the applications, we developed the 1PN version of the tensor virial theorem,
applicable for arbitrary self-gravitating systems. We found that the 1PN virial theorem differs from the Newtonian
one by the fact that the temporal variation of $\phi$ contributes to the variation of the inertia tensor.
However, for stationary systems we recover the same result as in Newtonian theory: the absolute value of the
total gravitational potential energy is two times the kinetic energy. The problem about the validity
of the virial theorem for collisional systems (which remains valid in the Newtonian theory) is an open question that we leave for future studies.

The applications we considered here were devoted specially to address the modeling of systems in galactic dynamics, thus we focused in equilibrium situations with rotational symmetry.
We developed a formalism to construct the 1PN version of Newtonian self-consistent models characterized by a stationary DF depending on the Jacobi's integral.
The general case is mathematically challenging as it involves a pair of coupled elliptic partial differential equations. However, we restricted our attention to the study of thin models with finite extent, the case in which the system reduces to a set of algebraic equations.
As an illustrative example
of the formalism developed here, we obtained the 1PN version of the second
MM disk. The contributions of relativistic corrections found are not significant
for $\lambda\sim10^{-3}$ or less, but they become prominent for larger values of $\lambda$ (provided that the 1PN approximation is still valid). We noted that the corrections grow with the
radial distance, which is surprising since one would expect major
corrections near the disk center, where the mass concentration is maximum.

The general formalism starts by implementing the Hunter's method in the 1PN scheme. Then, a stationary DF (solution of the 1PN CBE) is selected and introduced in the self-gravitation equations. Finally, by solving the resulting algebraic equalities,
we find the corresponding post-Newtonian fields, $\psi$ and $\xi_{i}$.
The new 1PN-corrected models obtained are also self-consistent,
since this feature is an inherent requirement of the formalism.
The method is applicable to self-gravitating thin disks with finite
extent provided that: (i) the system has symmetry
of reflection with respect to the $z=0$ plane; (ii) the Newtonian potential
 is separable in spheroidal oblate coordinates; (iii) The Newtonian DF is only function of $E$ and $L_{z}$.

It is worthwhile to point out the interesting work by Schenk, Shapiro and Teukolsky in 1999 \cite{shapiro}. In this paper the authors focused on the Kalnajs disk (the $m=1$ member of the MM family) and solved numerically the Einstein equations coupled to the relativistic CBE, obtaining the first self-consistent model of a rotating relativistic disk. It would be interesting to obtain the 1PN version of the same model by using the formalism developed in this paper in order to establish a comparison of the two approaches and estimate the effects of the higher order relativistic corrections.

As a final remark, it is important to mention that the problem about the stability of these models is an important subject that
will be addressed in future works. This issue brings several points to be investigated. One fundamental question is to establish whether the discoidal structures considered here are stable by themselves or if they need an additional component (a spheroidal halo, for example) in order to be stable. Another question is to determinate if the marginal stability of Newtonian disks improves or not with the introduction
of relativistic corrections. At the present we are performing a preliminary study about the stability of test particles around the solutions obtained, and their response to radial and vertical perturbations in order to compare the results with the Newtonian theory (see for example \cite{PRG,RLG}) and to provide a first test of stability. Then we expect to implement a more conclusive analysis regarding the Jeans-type instabilities by perturbing the DF's by themselves.

\section*{Acknowledgements}
This material is partially supported by the
National Science Foundation under Grant Number PHY-
0969020. The authors would like to thank Camilo Akimushkin and Guillermo Gonz\'alez for helpful discussions and
collaboration in the early stages of this project \cite{arcg}. The work of J.R.-C. is supported by Funda\c{c}\~ao de Amparo \`{a} Pesquisa do
Estado de S\~ao Paulo (FAPESP) grant 2009/16304-3. The work of C.A.A is supported by Mexico's National Council of Science and Technology
(CONACyT). J.F.P is grateful to the Texas Cosmology Center, which is supported by the College of Natural Sciences,
the Department of Astronomy at the University of Texas at Austin and the McDonald Observatory.

\appendix

\section{Rotation Curves in the 1PN Approximation}\label{Circular Velocity1PN}

The circular velocity is a widely used observable in galactic dynamics, so it is important to have in hand a
formula for the rotation curves in the 1PN approximation.
In order to do this, we have to study the circular motion of test particles in the equatorial plane.
At first, note that according to (\ref{ecmovimiento1PN}), a star moving around
an axisymmetric stationary thin disk with finite extension, obeys the following equations of motion:
\begin{eqnarray}
\ddot{R}&=&R\dot{\varphi}^{2} -\frac{\partial\phi}{\partial R}
\left(1 +\frac{1}{c^2}\left(4\phi-3\dot{R}^{2}+R^{2}\dot{\varphi}^{2}+\dot{z}^{2}\right)\right)\nonumber\\
&&+\frac{1}{c^2}\left(4\dot{R}\dot{z}\frac{\partial\phi}{\partial z}
-\frac{\partial\psi}{\partial R}+R\dot{\varphi}\frac{\partial\xi_{\varphi}}{\partial R}\right),\label{ecmotionR1PN}\\
&&\nonumber\\
\ddot{z} &=& -\frac{\partial\phi}{\partial z}
\left(1 +\frac{1}{c^2}\left(4\phi-3\dot{z}^{2}+R^{2}\dot{\varphi}^{2}+\dot{R}^{2}\right)\right)+\frac{1}{c^2}\left(4\dot{z}\dot{R}\frac{\partial\phi}{\partial R}
-\frac{\partial\psi}{\partial z}+R\dot{\varphi}\frac{\partial\xi_{\varphi}}{\partial z}\right),\label{ecmotionz1PN}\\
&&\nonumber\\
R\ddot{\varphi}&=&-2\dot{R}\dot{\varphi}+\frac{4R\dot{\varphi}}{c^{2}}
 \left(\dot{R}\frac{\partial\phi}{\partial R}
 +\dot{z}\frac{\partial\phi}{\partial z}\right)-\frac{\dot{R}}{c^{2}}\frac{\partial\xi_{\varphi}}{\partial R}-
 \frac{\dot{z}}{c^{2}}\frac{\partial\xi_{\varphi}}{\partial z},\label{ecmotionvarphi1PN}\nonumber\\
\end{eqnarray}
where the dot denotes derivation with respect to $t$.
In particular, equatorial circular orbits must satisfy the conditions $\dot{R}=\dot{z}=0$,
$\ddot{R}=\ddot{z}=0$ and $z=0$. In this case (\ref{ecmotionR1PN}) reduces to
\be
v_{\varphi}^{2}\left(1-\frac{R}{c^{2}}\frac{\partial\phi}{\partial R}\right)+
v_{\varphi}\frac{R}{c^{2}}\frac{\partial\xi_{\varphi}}{\partial R}-
R\frac{\partial\phi}{\partial R}-\frac{R}{c^{2}}\frac{\partial}{\partial R}(2\phi^{2}+\psi)=0,
\ee
which can be used to derive an expression for
$v_{\varphi}^{2}$ at 1PN order. In particular, note that we can use the Newtonian expression
$v_{\varphi}^2=R\partial\phi/\partial R$ each time that $v_{\varphi}$ is accompanied by an inverse power of $c$.
After some straightforward algebra we get
\be\label{rotlaw}
v_{\varphi}^{2}=
R\frac{\partial\phi}{\partial R}\left(1+\frac{4\phi}{c^{2}}+\frac{R}{c^{2}}
\frac{\partial\phi}{\partial R}\right)+\frac{R}{c^{2}}\left(\frac{\partial\psi}{\partial R}
-\sqrt{R\frac{\partial\phi}{\partial R}}\frac{\partial\xi_{\varphi}}{\partial R}\right),
\ee
where it is understood that all derivatives are evaluated at $z=0$.
There is a crucial difference between the above relation and the classical formula for the rotation
curves: in the Newtonian case $v_{\varphi}^{2}$ is linear in $\partial\phi/\partial R$, whereas
in the 1PN case, it depends in nonlinear terms involving $\phi$ and derivatives of the gravitational potentials. This
nonlinear dependence may be significant in some cases.

\section{Virial Theorem in the 1PN Approximation}\label{sec:ap-virial}

The virial theorem is an important general result
of the kinetic theory relating the contribution of kinetic and potential
energy to the temporal change of the inertia tensor associated to the system.
In order to obtain the 1PN version of the virial theorem we can start from
 the conservation laws
\bea
\frac{\partial T^{\mu\nu}}{\partial x^\mu}=-\Gamma^{\nu}_{\mu \lambda}T^{\mu \lambda}-\Gamma^{\mu}_{\mu \lambda}T^{\lambda \nu}.
\eea
At first order in $\bar{v}/c$, this expression reproduces
the Newtonian mass and momentum  conservation laws, which are given by
\bea
\label{masa}
\frac1c\frac{\partial \stackrel{0 \: \: \: \: }{T^{0 0}}}{\partial t}+
\frac{\partial \stackrel{1 \: \: \: \: }{T^{i 0}}}{\partial x^i}&=&0 \label{momento}\,, \\
\frac1c\frac{\partial \stackrel{1 \: \: \: \: }{T^{0 i}}}{\partial t}+
\frac{\partial \stackrel{2 \: \: \: \: }{T^{i j}}}{\partial x^j}&=&-\frac{1}{c^2}\frac{\partial \phi}{\partial x^i}\stackrel{0 \: \: \: \: }{T^{0 0}}\,.
\eea
The corresponding 1PN corrections of the above laws are obtained by taking into account Christoffel symbols at different orders \cite{WB}
\bea
\label{posmasa}
\frac1c\frac{\partial \stackrel{2 \: \: \: \: }{T^{0 0}}}{\partial t}+
\frac{\partial \stackrel{3 \: \: \: \: }{T^{i 0}}}{\partial x^i}&=&\frac{1}{c^3}\frac{\partial \phi}{\partial t}\stackrel{0 \: \: \: \: }{T^{00}}\,,
 \\
\label{posmomento}
\frac1c\frac{\partial \stackrel{3 \: \: \: \: }{T^{0 i}}}{\partial t}+\frac{\partial \stackrel{4 \: \: \: \: }{T^{i j}}}{\partial x^j}
&=&
-
\stackrel{4 \: \: \: \: }{\Gamma^{i}_{00}}\stackrel{0 \: \: \: \: }{T^{0 0}}
-
\stackrel{2 \: \: \: \: }{\Gamma^{i}_{00}}\stackrel{2 \: \: \: \: }{T^{0 0}}
-(2\stackrel{3 \: \: \: \: }{\Gamma^{i}_{0j}}+
\delta_{ij}\stackrel{3 \: \: \: \: }{\Gamma^{0}_{00}}+\delta_{ij}\stackrel{3 \: \: \: \: }{\Gamma^{k}_{0k}})\stackrel{1 \: \: \: \: }{T^{0 j}}
\nonumber \\ &&
-(\stackrel{2 \: \: \: \: }{\Gamma^{i}_{jk}}+
\delta_{ik}\stackrel{2 \: \: \: \: }{\Gamma^{0}_{0j}}+\delta_{ik}\stackrel{2 \: \: \: \: }{\Gamma^{l}_{lj}})\stackrel{2 \: \: \: \: }{T^{jk}}\,
\nonumber \\ &&
\nonumber \\
&=&
-
\Big[ \pxi \Big(\frac{2\phi^2}{c^4}+\frac{\psi}{c^4}\Big) + \frac{1}{c^4}\frac{\partial \xi_{i} }{\partial t} \Big]\stackrel{0 \: \: \: \: }{T^{0 0}}
-
\frac{1}{c^2}\frac{\partial \phi}{\partial x^i}\stackrel{2 \: \: \: \: }{T^{0 0}}\nonumber \\ &&
-\frac{1}{c^3}\Big(\frac{\partial \xi_i}{\partial x^j}-\frac{\partial \xi_j}{\partial x^i}
-4\delta_{ij} \frac{\partial \phi}{\partial t}\Big)    \stackrel{1 \: \: \: \: }{T^{0 j}}
-\frac{1}{c^2}\Big( \delta_{jk}\frac{\partial \phi}{\partial x^i}-4\delta_{ik}\frac{\partial \phi}{\partial x^j}\Big)\stackrel{2 \: \: \: \: }{T^{jk}}\,.
\eea
On the other hand, the energy-momentum tensor, given in (\ref{T1PN-1}), can be split in two first-order contributions:
\bea
T^{\mu \nu}=\int\stackrel{0}{F} V^{\mu}V^{\nu}d^{3}v +
\int\Big[\stackrel{0}{F}\Big(\frac{3{\bf v}^2}{c^2}-\frac{6\phi}{c^2}\Big)+\stackrel{2}{F}\Big] V^{\mu}V^{\nu}d^{3}v
\eea
where $V^\mu=(c,{\bf v})$. After some definitions
\bea
\stackrel{0}{\rho}\, \equiv \int \stackrel{0}{F}d^{3}v, \qquad \stackrel{2}{\rho}\, \equiv \int \stackrel{2}{F}d^{3}v,\quad\textrm{and}\quad
\stackrel{2}{\tilde{\rho}}\, \equiv \,\stackrel{2}{\rho}+\stackrel{0}{\rho}\Big(\frac{3 \overline{{\bf v}^2} }{c^2}-\frac{6\phi}{c^2}\Big),
\eea
and using the different probability densities to compute expectation values,
\bea
\overline{\cal A}&=&(\stackrel{0}{\rho})^{-1}\int\stackrel{0}{F}{\cal A}d^3v\,,
\nonumber \\
\stackrel{2}{\overline{ \cal A}}&=&({\stackrel{2}{\tilde{\rho}}})^{-1}
\int\Big[\stackrel{0}{F}\Big(\frac{3{\bf v}^2}{c^2}-\frac{6\phi}{c^2}\Big)+\stackrel{2}{F}\Big]{\cal A}d^3v\,,
\eea
we get the following simplified expressions for the momentum-energy components at different orders:
\bea
&&\stackrel{0 \: \: \: \: }{T^{0 0}}=c^2\stackrel{0}{\rho}\,,
\qquad \stackrel{2 \: \: \: \: }{T^{0 0}}=c^2\stackrel{2}{\tilde{\rho}}\,, \qquad \stackrel{1 \: \: \: \: }{T^{0
 i}}=c\stackrel{0}{\rho}{\overline{v^i}}\,,\\
&&\stackrel{3 \: \: \: \: }{T^{0 i}}=c\stackrel{2}{\rho}\stackrel{2}{{\overline{v^i}}}\,,\qquad
\stackrel{2 \: \: \: \: }{T^{i j}}=\stackrel{0}{\rho}{\overline{v^iv^j }}\,, \qquad
\stackrel{4 \: \: \: \: }{T^{i j}}=\stackrel{2}{\tilde{\rho}}\stackrel{2}{{\overline{v^iv^j}}}\,.
\eea
Then, equations (\ref{posmasa})-(\ref{posmomento}), can be written as
\bea
\label{posmasa2}
\frac{\partial (\stackrel{2}{\tilde{\rho}})}{\partial t}+
\frac{\partial (\stackrel{2}{\tilde{\rho}}\stackrel{2}{{\overline{v^i}}})}{\partial x^i}&=&\frac{\stackrel{0}{\rho}}{c^2}
\frac{\partial \phi}{\partial t}\,,
\\
\label{posmomento2}
\frac{\partial (\stackrel{2}{\tilde{\rho}}\stackrel{2}{{\overline{v^i}}})}{\partial t}
+
\frac{\partial (\stackrel{2}{\tilde{\rho}}\stackrel{2}{\overline{v^iv^j}})}{\partial x^j}
&=&
-
\stackrel{0}{\rho}\Big[ \pxi \Big(\frac{2\phi^2}{c^2}+\frac{\psi}{c^2}\Big) + \frac{1}{c^2}\frac{\partial \xi_i}{\partial t}\Big]
-
\stackrel{2}{\tilde{\rho}}\frac{\partial \phi}{\partial x^i}\nonumber \\
&&
-
\stackrel{0}{\rho}\frac{\overline{v^j}}{c^2}\Big(\frac{\partial \xi_i}{\partial x^j}
-\frac{\partial \xi_j}{\partial x^i}-4\delta_{ij} \frac{\partial \phi}{\partial t}\Big)
-
\stackrel{0}{\rho}\frac{\overline{v^jv^k}}{c^2}\Big( \delta_{jk}\frac{\partial \phi}{\partial x^i}
-4\delta_{ik}\frac{\partial \phi}{\partial x^j}\Big).
\eea
Now we can compute the symmetrized integral of the product $x^k/2$ times the above equation:
\bea
\frac12\frac{d}{dt}\int d^3x \stackrel{2}{\tilde{\rho}}
(x^k\stackrel{2}{\overline{v^i}}+x^i\stackrel{2}{\overline{v^k}})=2\stackrel{2 \: \: \:}{K^{ik}}+\stackrel{2 \: \: \:}{W^{ik}}
\eea
where
\bea
\stackrel{2 \: \: \:}{W^{ik}}&=&
-
\frac12\int d^3x \stackrel{0}{\rho} x^k \left[ \pxi \Big(\frac{2\phi^2}{c^2}+ \frac{\psi}{c^2}\Big) + \frac{1}{c^2}\frac{\partial \xi_i}{\partial t}
+
\frac{\overline{v^j}}{c^2}\Big(\frac{\partial \xi_i}{\partial x^j}
-\frac{\partial \xi_j}{\partial x^i}-4\delta_{ij} \frac{\partial \phi}{\partial t}\Big)\right.\nonumber\\
&&\quad\quad\quad\quad\quad\quad\quad\,+
\left.\frac{\overline{v^jv^l}}{c^2}\Big( \delta_{jl}\frac{\partial \phi}{\partial x^i}-4\delta_{il}\frac{\partial \phi}{\partial x^j}\Big)\right] \nonumber \\
&&-
\frac12\int d^3x \stackrel{0}{\rho} x^i \left[ \frac{\partial}{\partial x^k} \Big(\frac{2\phi^2}{c^2}+ \frac{\psi}{c^2}\Big)
 + \frac{1}{c^2}\frac{\partial \xi_k}{\partial t}
+
\frac{\overline{v^j}}{c^2}\Big(\frac{\partial \xi_k}{\partial x^j}-\frac{\partial \xi_j}{\partial x^k}-4\delta_{kj}
\frac{\partial \phi}{\partial t}\Big)\right.\nonumber\\
&&\quad\quad\quad\quad\quad\quad\quad\,+
\left.\frac{\overline{v^jv^l}}{c^2}\Big( \delta_{jl}\frac{\partial \phi}{\partial x^k}-4\delta_{kl}\frac{\partial \phi}{\partial x^j}\Big)\right] \nonumber \\
&&-\frac12\int d^3x \stackrel{2}{\tilde{\rho}}\left( x^k\frac{\partial \phi}{\partial x^i}+x^i\frac{\partial \phi}{\partial x^k}\right)\,,\\
\stackrel{2 \: \: \:}{K^{ik}}&=&\frac12\int d^3x \stackrel{2}{\tilde{\rho}}{\overline{v^jv^k}}\,.\label{potential-energy1}
\eea

By defining
\bea
\stackrel{2 \: \: \:}{I^{ik}}= \int d^3x \stackrel{2}{\tilde{\rho}}x^ix^k,
\eea
we can write the time derivative of the equation (\ref{posmasa}) as
\bea
\frac{d \stackrel{2 \: \: \:}{I^{ik}} }{dt}=\int d^3x
\frac{\partial \stackrel{2}{\tilde{\rho}} }{\partial t}x^ix^k=
\int d^3x \stackrel{2}{\tilde{\rho}}
(x^k\stackrel{2}{\overline{v^i}}+x^i\stackrel{2}{\overline{v^k}})
+
\int d^3x \frac{\stackrel{0}{\rho}}{c^2}\frac{\partial \phi }{\partial t}x^ix^k
\eea
and its second time derivative
\bea
\frac{d^2 \stackrel{2 \: \: \:}{I^{ik}} }{dt^2}=\frac{d}{dt}\int d^3x \stackrel{2}{\tilde{\rho}}
(x^k\stackrel{2}{\overline{v^i}}+x^i\stackrel{2}{\overline{v^k}})
+ \frac{d}{dt}\int d^3x \frac{\stackrel{0}{\rho}}{c^2}\frac{\partial \phi }{\partial t}x^ix^k\,.
\eea
Putting all together we obtain the tensor post-Newtonian virial theorem
\bea
\frac{d^2 \stackrel{2 \: \: \:}{I^{ik}} }{dt^2}=2\stackrel{2 \: \: \:}{K^{ik}}
+\stackrel{2 \: \: \:}{W^{ik}}+ \frac12\frac{d}{d t}\int d^3x \frac{\stackrel{0}{\rho}}{c^2}\frac{\partial \phi }{\partial t}x^ix^k\,,
\eea
and taking the trace, we obtain the scalar post-Newtonian virial theorem
\bea
\frac{d^2 \stackrel{2 }{I} }{dt^2}=2\stackrel{2}{K}+\stackrel{2}{W}+
 \frac12\frac{d}{d t}\int d^3x \frac{\stackrel{0}{\rho}}{c^2}\frac{\partial \phi }{\partial t}{\bf x}^2\,,
\eea
where
\bea
\stackrel{2 }{O}=\mathrm{trace}(\stackrel{2 \: \: \:}{O^{ik}})\,.
\eea

In summary, we can collect all the above results and state that if we define a moment of inertia tensor,
\bea
I^{ik}= \int \int \left(1+\frac{3\overline{v^{2}}}{c^{2}}-\frac{6\phi}{c^{2}}\right)x^ix^k F d^3v d^3x,
\eea
a kinetic energy tensor,
\bea
K^{ik}= \frac{1}{2}\int \int \left(1+\frac{3\overline{v^{2}}}{c^{2}}-\frac{6\phi}{c^{2}}\right)\overline{v^iv^k} F d^3v d^3x,
\eea
and a potential energy tensor
\bea
W^{ik}= \stackrel{0\: \: \: \:}{W^{ik}}+\stackrel{2\: \: \: \:}{W^{ik}},
\eea
where $\stackrel{0\: \: \: \:}{W^{ik}}$ is the Newtonian potential energy tensor and $\stackrel{2\: \: \: \:}{W^{ik}}$
is given by (\ref{potential-energy1}), then they satisfy the following relation
\bea
\frac{d^2 I^{ik} }{dt^2}=2K^{ik}
+W^{ik}+ \frac12\frac{d}{d t}\int d^3x \frac{\stackrel{0}{\rho}}{c^2}\frac{\partial \phi }{\partial t}x^ix^k\,,
\eea
which can be enunciated as the 1PN tensor virial theorem. The scalar virial equation is obtained by
taking the trace of the above relation:
$$
\frac{d^2 I}{dt^2}=2K
+W+ \frac12\frac{d}{d t}\int d^3x \frac{\stackrel{0}{\rho}}{c^2}\frac{\partial \phi }{\partial t}\mathbf{x}^2.
$$

\section{Jacobi's Integral in the 1PN Approximation}\label{ap1}

For the purposes of the present paper it is important to have in hand an expression of the Jacobi's integral
at 1PN order, in order to verify that it has the same form as in Newtonian theory, i.e. $ E_{J}=E-\Omega L_{z}$.

In a rotating reference frame with angular velocity $\mathbf{\Omega}$, velocities are related through
\begin{equation}\label{trans-velocidad}
    \mathbf{\tilde{v}}=\mathbf{v}-\mathbf{\Omega}\times \mathbf{x},
\end{equation}
where $\mathbf{\tilde{v}}$ and $\mathbf{v}$ are the
velocity measured from the rotating and inertial frame, respectively.
 From this relation, and the 1PN equations of motion it is possible to derive a 1PN corrected version for $E_{J}$ \cite{nelson}:
\begin{eqnarray}\label{Ej}
    E_{J}&=&\frac{\tilde{v}^{2}}{2}+\phi-\frac{1}{2}(\mathbf{\Omega}\times
    \mathbf{x})^{2}+\frac{1}{c^{2}}\left[\frac{3\tilde{v}^{4}}{8}+\frac{\phi^{2}}{2}+\psi
    -\frac{3\phi\tilde{v}^{2}}{2}-(\mathbf{\Omega}\times \mathbf{x})\cdot\mbox{\boldmath$\xi$}+\frac{3\phi}{2}(\mathbf{\Omega}\times \mathbf{x})^{2}
    \right.\nonumber\\&&\nonumber\\ && \left.
    \qquad\qquad\quad-\frac{1}{8}(\mathbf{\Omega}\times \mathbf{x})^{4}+\frac{\tilde{v}^{2}}{4}(\mathbf{\Omega}\times \mathbf{x})^{2}
    +\mathbf{\tilde{v}}\cdot(\mathbf{\Omega}\times \mathbf{x})\tilde{v}^{2}+
    \frac{1}{2}(\mathbf{\tilde{v}}\cdot(\mathbf{\Omega}\times \mathbf{x}))^{2}\right].
\end{eqnarray}
In particular, if $\mathbf{\Omega}=\Omega \mathbf{\hat{e}_{z}}$  and $\mathbf{x}=R\mathbf{\hat{e}_{R}}$,
the above formula reduces to
\begin{eqnarray}
    E_{J}=\frac{\tilde{v}^{2}}{2}+\phi-\frac{\Omega^{2} R^{2}}{2}+\frac{1}{2 c^{2}}\left[\frac{3\tilde{v}^{4}}{4}+\phi^{2} + 2\psi
    -3\phi\tilde{v}^{2}-2\Omega R \xi_{\varphi}+3\phi\Omega^{2} R^{2}
    \right.\nonumber\\ \nonumber\\ \left.\quad-\frac{\Omega^{4} R^{4}}{4}+\frac{\tilde{v}^{2}\Omega^{2} R^{2}}{2}
    +2\Omega R\tilde{v}_{\varphi}\tilde{v}^{2}+
    \tilde{v}_{\varphi}^{2}\Omega^{2} R^{2}\right].\label{Ej2}
\end{eqnarray}
Now, by replacing $\tilde{v}_{R}=v_{R}$, $\tilde{v}_{z}=v_{z}$, and $\tilde{v}_{\varphi}=v_{\varphi}-\Omega R$
in (\ref{Ej2}), we can write
\begin{equation}\label{rel-Ej-E}
    E_{J}=E-\Omega L_{z},
\end{equation}
where $E$ and $L_{z}$ are given by (\ref{ec_ene}) and (\ref{lz_pn}), respectively. Thus we see that the
total energy associated to a test particle, measured from a rotating frame, has the same form as in the
Newtonian theory.

\section{Derivation of the Constants $B_{2n}$ and $C_{2n}$}\label{ap2}

For the DF given in (\ref{def-k-omega}) and for any function $S(R,v_{R},v_{\varphi})$, we can write
\begin{equation}\label{prom-f0}
    \int S fd^{2}v=2^{1/4}k\int_{\Omega R-\mu}^{\Omega R+\mu}dv_{\varphi}\int_{-\nu}^{\nu}
    \frac{S(R,v_{R},v_{\varphi})dv_{R}}{(\nu^{2}-v_{R}^{2})^{1/4}}
\end{equation}
where
\begin{equation}\label{mu-nu}
    \mu=\sqrt{\frac{45\pi GM}{64a}}\eta_{*}^{2},\qquad \nu=\sqrt{\mu^{2}-(v_{\varphi}-\Omega R)^{2}}.
\end{equation}
In particular, by setting $S=v_{\varphi}$ in (\ref{prom-f0}) and introducing
the result in (\ref{selfgrav-3-morgan}) we obtain
$$
    \sum_{n = 0}^{\infty} \tilde{C}_{2n} P_{2n}(\eta_{*})=\frac{5}{8}
    \sqrt{\frac{15GM^{3}}{2\pi a^{5}}}\sqrt{1-\eta_{*}^{2}}\eta_{*}^{4},
$$
from which $\tilde{C}_{2n}$ can be found by using the orthogonalization
condition $\int_{-1}^{1}P_{m}(x)P_{l}(x)dx=2\delta_{lm}/(2l+1)$. We find
\begin{equation}\label{C2ntilde-1}
    \tilde{C}_{2n}=(4n+1)\frac{5}{8}
    \sqrt{\frac{15GM^{3}}{2\pi a^{5}}}\int_{0}^{1}\sqrt{1-x^{2}}x^{4}P_{2n}(x)dx.
\end{equation}
The above integral can be solved by setting $x=\cos\theta$ and using
the following identity:
\begin{equation}
    \int_{0}^{\pi/2}(\sin\theta)^{m} P_{2n}(\cos\theta)d\theta=
    \frac{\pi\left[\Gamma(\frac{m+1}{2})\right]^{2}}{2\Gamma(1+n+\frac{m}{2})
    \Gamma(\frac{1}{2}+\frac{m}{2}-n)\Gamma(1+n)\Gamma(\frac{1}{2}-n)}.\label{abramowitz}
\end{equation}
After some calculations, we get
\begin{equation}
    \int_{0}^{1}\sqrt{1-x^{2}}x^{4}P_{2n}(x)dx=
    \frac{\pi^{2}(45-48n-92n^{2}+16n^{3}+16n^{4})}{128\Gamma(\frac{1}{2}-n)
    \Gamma(\frac{7}{2}-n)\Gamma(1+n)\Gamma(4+n)}.\label{int-C2ntilde}
\end{equation}
Plugging the above result in (\ref{C2ntilde-1}) we obtain the relation (\ref{C2n}).

Now, performing the same procedure on equation (\ref{selfgrav-2-morgan}) and obtain
$$
\pi^{2}G a\sum_{n = 0}^{\infty} \tilde{B}_{2n} P_{2n}(\eta_{*})=
\frac{75\pi^{2}}{7168}\left(\frac{GM}{a}\right)^{2} H(\eta_{*})+\frac{32\psi}{3}+
\sqrt{\frac{160\pi GM}{3a}(1-\eta_{*}^{2})}\xi_{\varphi}
$$
where
\begin{equation}\label{H}
    H(\eta_{*})=375\eta_{*}^{8}-154\eta_{*}^{6}+392\eta_{*}^{4}+14\eta_{*}^{2}-217.
\end{equation}
Again, we use the orthogonality of Legendre polynomials to provide an explicit relation for $\tilde{B}_{2n}$.
In the calculations we also find useful (i) the formula
$$
\int_{-1}^{1}x^{2m}P_{2n}(x)dx=\frac{\sqrt{\pi}2^{-2m}\Gamma(1+2m)}{\Gamma(1+m-n)\Gamma(\frac{3}{2}+m+n)},
$$
and (ii) the expression
\begin{eqnarray}
    \int_{-1}^{1}\sqrt{1-x^{2}}P_{2n}(x)P_{2m}(x)dx= \sum_{k=0}^{m}\sum_{j=0}^{m-k}\frac{\pi(4m-2k)!
    (m-k)!}{ 2^{2m}k!(2m-k)!j!(m-k-j)!(2m-2k)!}\nonumber\\
     \times\frac{(-1)^{k+j}\left[\Gamma(j+\frac{3}{2})\right]^{2}}{\Gamma(2+j+n)\Gamma(j-n+\frac{3}{2})\Gamma(1+n)\Gamma(\frac{1}{2}-n)},
    \label{G2n2m}
\end{eqnarray}
which we derived from (\ref{abramowitz}) and the series representation of the Legendre polynomials.
After some calculations we get
\begin{equation}
\frac{64}{3} B_{2n}q_{2n}(0)+2\pi^{2} a G \tilde{B}_{2n}=
\left(\frac{GM}{a}\right)^{2}\left[I_{2n}-\sum_{m=0}^{\infty} \Pi_{2n,2m}\right],\label{ecua}
\end{equation}
where
\begin{equation}\label{I2n}
    I_{2n}=-\frac{75\pi^{5/2}}{16384}\frac{(4n+1)\sum_{k=0}^{8}a_{k}n^{k}}{\Gamma(5-n)\Gamma(\frac{11}{2}+n)},
\end{equation}
with
\begin{eqnarray}
a_{0}=370224\qquad a_{1}=-337854\qquad a_{2}=-628841\nonumber\\
a_{3}=185300\qquad a_{4}=174491\qquad\:\:\:\: a_{5}=-25768\:\:\nonumber\\
a_{6}=16600\qquad\:\: a_{7}=992\qquad\qquad\: a_{8}=496\qquad\:\label{constantes a}
\end{eqnarray}
and
\begin{equation}\label{Pi2n}
    \Pi_{2n,2m}=\frac{25\pi^{3}}{64}\int_{-1}^{1}\sqrt{1-x^{2}}P_{2n}(x)P_{2m}(x)dx.
\end{equation}
From (\ref{ecua}), it is straightforward to derive (\ref{B2n}).

\end{document}